\documentclass[acmsmall]{acmart}
\AtBeginDocument{%
	}

\setcopyright{acmcopyright}
\copyrightyear{2024}
\acmYear{2024}
\acmDOI{XXXXXXX.XXXXXXX}

\acmISBN{978-1-4503-XXXX-X/18/06}

\usepackage{enumitem}
\usepackage{hyperref}
\usepackage[linesnumbered,lined,ruled,commentsnumbered]{algorithm2e}
\usepackage{listings, xcolor, soul} 
\usepackage{mathrsfs}
\usepackage{amsmath}
\usepackage{multirow}
\usepackage{diagbox}
\usepackage{booktabs}
\usepackage{color, colortbl}
\usepackage{array}
\usepackage{graphicx}
\usepackage{caption}
\let\origthelstnumber\thelstnumber
\makeatletter
\newcommand*\Suppressnumber{%
	\lst@AddToHook{OnNewLine}{%
		\let\thelstnumber\relax%
		\advance\c@lstnumber-\@ne\relax%
	}%
}

\newcommand*\Reactivatenumber[1]{%
	\setcounter{lstnumber}{\numexpr#1-1\relax}
	\lst@AddToHook{OnNewLine}{%
		\let\thelstnumber\origthelstnumber%
		\refstepcounter{lstnumber}
	}%
}

\makeatother

\definecolor{Gray}{gray}{0.6}
\definecolor{LightGray}{gray}{0.9}
\definecolor{bananayellow}{rgb}{1.0, 0.88, 0.5}

\newcommand{\toolname}{\textsc{TripRL}\xspace}
\newcolumntype{a}{>{\columncolor{Gray}}c}
\newcolumntype{b}{>{\columncolor{LightGray}}c}

\newboolean{showcomments}
\setboolean{showcomments}{true}
\ifthenelse{\boolean{showcomments}}
{
	\definecolor{myyellow}{RGB}{255, 228, 26}
	\definecolor{myblue}{RGB}{50, 50, 220}
	\newcommand{\nb}[2]{
		{\sf
			{\color{myblue}\fontsize{7pt}{8pt}\selectfont\textbf{#2}}%
		}%
	}
}
{
	\newcommand{\nb}[2]{}
}

\newcommand{\header}[1]{\noindent\textbf{#1}}
\SetKwComment{Comment}{/* }{ */}

\begin{document}
	
	\title{Scalable Similarity-Aware Test Suite Minimization with Reinforcement Learning}
	
	\author{Sijia Gu}
	\affiliation{%
		\department{Electrical and Computer Engineering}
		\institution{University of British Columbia}
		\city{Vancouver}
		\state{BC}
		\country{Canada}}
	\email{sijiagu@ece.ubc.ca}
	
	\author{Ali Mesbah}
	\affiliation{%
		\department{Electrical and Computer Engineering}
		\institution{University of British Columbia}
		\city{Vancouver}
		\state{BC}
		\country{Canada}}
	\email{amesbah@ece.ubc.ca}
	\begin{abstract}	
		The Multi-Criteria Test Suite Minimization (MCTSM) problem aims to remove redundant test cases, guided by adequacy criteria such as code coverage or fault detection capability. However, current techniques either exhibit a high loss of fault detection ability or face scalability challenges due to the NP-hard nature of the problem, which limits their practical utility. We propose \toolname, a novel technique that integrates traditional criteria such as statement coverage and fault detection ability with test coverage similarity into an Integer Linear Program (ILP), to produce a diverse reduced test suite with high test effectiveness. \toolname leverages bipartite graph representation and its embedding for concise ILP formulation and combines ILP with effective reinforcement learning (RL) training. This combination renders large-scale test suite minimization more scalable and enhances test effectiveness. Our empirical evaluations demonstrate that \toolname's runtime scales linearly with the magnitude of the MCTSM problem. Notably, for large test suites from the Defects4j dataset where existing approaches fail to provide solutions within a reasonable time frame, our technique consistently delivers solutions in less than 47 minutes. The reduced test suites produced by \toolname also maintain the original statement coverage and fault detection ability while having a higher potential to detect unknown faults.
	\end{abstract}

	
	\maketitle
	
	\section{Introduction}\label{intro}
	Testing is fundamental in software development, safeguarding the integrity and functionality of software systems. However, as production code continuously evolves, the accompanying test suites often inflate with redundant test cases, which may not always enhance code coverage or efficacy \cite{bavota2012empirical, hauptmann2015generating}. This in turn exacerbates test maintenance costs and efforts. Consequently, test suite minimization techniques have emerged \cite{vahabzadeh2018fine, lin2018nemo, khan2018systematic}, aiming to prune redundant test cases, ultimately curbing testing overheads. They operate on metrics such as code coverage, fault detection capability, and test runtime to guide the pruning process. Minimization techniques that follow a sole criterion, such as statement coverage, can sometimes undermine other criteria, such as fault detection capability \cite{rothermel1998empirical}. Addressing this, multi-criteria test suite minimization (MCTSM) aims to harmonize various minimization criteria. For example, a minimized test suite should retain statement coverage while maximizing fault detection capability.
	
	Current MCTSM solutions \cite{lin2018nemo, hsu2009mints, hao2012demand, black2004bi,  xue2020multi, ozener2020effective} capitalize on Integer Linear Programming (ILP) for problem formulation. Fundamentally, an ILP problem seeks to optimize a linear objective function with binary variables constrained by linear conditions \cite{wolsey2020integer}. Prevailing techniques formulate MCTSM problems of assorted criteria into individualized ILP models, which ILP solvers such as CPLEX~\cite{cplex} and Gurobi~\cite{gurobi} then solve. Nevertheless, the inherent complexity of MCTSM, being NP-hard \cite{hsu2009mints, black2004bi}, limits these techniques mainly to modestly-sized test suites with only hundreds of test cases \cite{lin2018nemo, xue2020multi,ozener2020effective}. As the criteria become more complex and the size of test suites increases, computational demands soar, rendering it impractical for voluminous test suites \cite{lin2018nemo}. In practice, industrial test suites \cite{petrovic2021does} are often much larger in scale than these techniques can handle effectively. 
	
	In contrast to MCTSM, which extracts coverage information between production and test code, black-box test minimization techniques such as FAST-R \cite{cruciani2019scalable} and ATM \cite{pan2023atm} rely solely on test code to prioritize computational speed. The empirical results of FAST-R \cite{cruciani2019scalable} indicate that this approach may compromise the fault detection rate of the test suite. Additionally, empirical evidence from ATM \cite{pan2023atm} reveals that these techniques still face scalability challenges when handling large test suites with nearly 4,000 test cases. 
	
	In this work, we introduce an innovative technique called \toolname that incorporates pairwise similarity into the ILP formulation for the MCTSM problem, focusing on two pivotal criteria: statement coverage and fault detection efficacy. This approach aims to produce a more diverse reduced test suite for improved test effectiveness. Our insight is that by adopting  Reinforcement Learning (RL) to solve the unique ILP model, we can enhance time efficiency and scalability for large test suites that would otherwise be unsolvable by a pure ILP approach. Our technique begins by formulating the MCTSM problem into an ILP model, constrained by statement coverage and fault detection ability, with the objective of minimizing both test suite size and coverage-based similarity. We then employ an RL agent, trained to refine the iterative process, which learns from a bipartite graph-represented environment mirroring our MCTSM problem. Through interactions with this environment, the RL agent produces a reduced set of test cases feasible within the ILP constraints and aims to minimize the ILP objective function. By combining RL for intelligent test suite reduction with our novel ILP formulation, our approach efficiently addresses the MCTSM problem and scales to larger test suites with improved test effectiveness than existing test minimization approaches~\cite{lin2018nemo, hsu2009mints, hao2012demand, black2004bi,  xue2020multi, ozener2020effective, cruciani2019scalable, pan2023atm}.
	
	Our work makes the following contributions:
	
	\begin{itemize}[itemsep=1pt, topsep=2pt, partopsep=0pt]
		\item{An approach, called \toolname, which leverages the strengths of both reinforcement learning and integer linear programming for tackling the MCTSM problem. We are the first to employ reinforcement learning in test minimization, to our knowledge. \toolname is designed for scalability and can reduce large test suites effectively.} 
		
		\item{A novel ILP formulation of the MCTSM problem by incorporating pairwise coverage-based similarity between tests to yield a diverse test suite with a higher potential to detect defects.} 
		
		\item{The first formulation of the MCTSM problem as a bipartite graph to the best of our knowledge, enabling us to leverage efficient bipartite graph embeddings in the reinforcement learning process.}
		
		\item{A comprehensive empirical evaluation, encompassing real-world open-source programs from the Defects4J~\cite{just2014defects4j} benchmark.}
	\end{itemize}
	
	Our results demonstrate that the time required by \toolname to minimize tests scales linearly with the size of the MCTSM problem. For large test suites where ILP solvers fail to produce solutions within a reasonable time (i.e., several hours), \toolname consistently delivers solutions in less than an hour. Moreover, \toolname outperforms existing state-of-the-art  techniques~\cite{pan2023atm, lin2018nemo} in statement coverage and achieves an increase of up to 33.5\% in fault-finding effectiveness.
	
	\section{Background and Challenges}	\label{challenges} 
	
	The Multi-Criteria Test Suite Minimization (MCTSM) problem optimizes the selection of a reduced test suite from a set of alternatives, meeting specific criteria from the original suite. Common criteria include reduced test suite size, statement coverage, fault detection capability, and execution time. For example, in Table \ref{tab:example}, the test suite \( T \) includes three test cases \(\{t_1, t_2, t_3\}\). Criteria such as statement coverage and detected faults are collected for each test case, represented by 0–1 matrices. For instance, \( t_1 \) covers statement \( s_1 \) and detects fault \( f_4 \), so elements \( s_1 \) and \( f_4 \) in row \( t_1 \) are marked as 1, while others are 0.
	
	\header{TSM Instance.}	In this paper, we refer to the original test suite with its criteria data as a \textit{TSM instance}, used to generate the reduced test suite (e.g., Table \ref{tab:example}). For the MCTSM problem, the main task is to decide whether to include or exclude each test case in the reduced suite, using binary decision variables \( t_i \). Here, \( t_i = 1 \) if the \( i \)-th test case is included, and \( t_i = 0 \) otherwise. The element \( s_{ip} \) in the statement matrix shows if the \( i \)-th test case covers statement \( s_p \), and \( f_{ik} \) in the fault matrix shows if it reveals fault \( f_k \). Integer Programming (IP)~\cite{wolsey2020integer} is a suitable method for modeling MCTSM problems.
	
	\begin{table}
		\caption{Statement coverage and fault detection data for sample test suite $T=\{t_1, t_2, t_3\}$}
		\label{tab:example}
		\small
		\begin{center}
			\begin{tabular}{|cc|ccc|cccc|} \hline
				\multicolumn{2}{|c|}{\multirow{2}{*}{\diagbox{Test Cases}{Criteria}}}& \multicolumn{3}{c|}{\small{Statement}} &\multicolumn{4}{c|}{\small{Fault}}  \\ 
				\multicolumn{2}{|c|}{ }&$s_1$&$s_2$&$s_3$&$f_1$&$f_2$&$f_3$&$f_4$\\
				\hline
				\multicolumn{2}{|c|}{$t_1$}&1&0&0&0&0&0&1 \\
				\hline
				\multicolumn{2}{|c|}{$t_2$}&0&1&1&1&1&1&0 \\
				\hline
				\multicolumn{2}{|c|}{$t_3$}&1&0&1&1&1&1&0 \\
				\hline
			\end{tabular}
		\end{center}
	\end{table}
	
	\header{The Classical Bi-criteria Model.} According to the MINTS~\cite{hsu2009mints} framework, criteria in MCTSM can be classified as either \textit{relative} or \textit{absolute} based on problem specifications. An \textit{absolute} criterion acts as a constraint in the minimization problem, while a \textit{relative} criterion defines an objective. In the classical bi-criteria model, statement coverage is the only \textit{absolute} criterion, serving as the constraint to ensure equivalent coverage to the original test suite after minimization. The reduced test suite's size and the number of uncovered faults are the \textit{relative} criteria, i.e., the objectives. The set of equations \ref{ilp_1}--\ref{ilp_3} encapsulate this model~\cite{black2004bi, hsu2009mints}.
	{\small
		\begin{align}
		\textrm{Minimize  } &\sum_{1 \leq i \leq |T|} t_{i} - \sum_{1 \leq i \leq |T|} w_o(t_i)t_{i}\label{ilp_1}\\
		\textrm{Subject To:  } 
		&\sum_{1 \leq i \leq |T|} s_{ip}t_i \geq 1, \quad 1 \leq p \leq |S| \label{ilp_2}\\
		&t_{i}\in \{0,1\}, \quad 1\leq i\leq |T|\label{ilp_3}
		\end{align}
	}%
	The statement coverage matrix is \( (s_{ip})_{1 \leq i \leq |T|, 1 \leq p \leq |S|} \), and the fault matrix is \( (f_{ik})_{1 \leq i \leq |T|, 1 \leq k \leq |F|} \), for \( |S| \) code statements and \( |F| \) faults. A test suite \( T \) with \( |T| \) test cases is represented by binary variables \( t_i \) (\( 1 \leq i \leq |T| \)), as shown in equation \ref{ilp_3}. Equation \ref{ilp_2} maintains the same statement coverage before and after minimization, ensuring each statement \(s_p\) is covered at least once. Equation \ref{ilp_1} stipulates the objectives to minimize the test suite size (\(\sum_{i=1}^{|T|} t_{i}\)) and maximize the number of detected faults (\(- \sum_{i=1}^{|T|} w_o(t_i)t_{i}\)). The weight function for fault detection capability is defined as the ratio of faults identified by the test case to the total known faults: \(w_o(t_i) = (\sum_{k=1}^{|F|}f_{ik})/|F|\). The optimal solution for the TSM instance in Table \ref{tab:example} is \(\{ t_2, t_3 \}\), minimizing the objective function in Equation \ref{ilp_1} as per the bi-criteria ILP model.
	
	\header{Nonlinear Problem Formulation.} Lin et al. \cite{lin2018nemo} identified the sub-optimal fault detection performance of the classical bi-criteria model \cite{hao2012demand, hsu2009mints}. They noted that the classical model often misses distinct faults. For example, the minimized solution \(\{t_2, t_3\}\) from the bi-criteria model for Table \ref{tab:example} misses fault \(f_4\), whereas the optimal solution \(\{t_1, t_2\}\) detects all faults. They introduced Nemo \cite{lin2018nemo}, a model with nonlinear weight function \( \widetilde{w}_o(t_i) \) to maximize distinct fault detection. However, the nonlinear approach increased the scalability challenges. Their evaluations, with test suites up to 746 test cases, often failed to conclude within an eight-hour timeframe, experiencing timeouts as noted by Xue et al. \cite{xue2020multi}. They later linearized the approach with auxiliary decision variables to make it solvable by ILP solvers. Despite aiming to maximize distinct faults, Nemo still suffers from significant fault detection loss.

	\header{Challenges and limitations.} Existing methodologies for MCTSM problems \cite{ozener2020effective, xue2020multi, lin2018nemo, hao2012demand, hsu2009mints, black2004bi, baller2014multi} typically revolve around formulations using Integer Linear/Non-Linear Programming and rely on commercial ILP solvers like CPLEX \cite{cplex} or Gurobi \cite{gurobi}. However, their effective application is often limited to small-to-medium-sized test suites due to scalability challenges. The MCTSM problem, classified as NP-hard \cite{hsu2009mints, wolsey2020integer}, poses difficulty in finding optimal solutions for large instances within a reasonable timeframe. Additionally, approaches like MINTS and Nemo often overlook enhancing the test suite's capability to detect unknown faults in the future. Even if the minimized test suite covers all statements and known faults, its test cases might belong to the same feature, potentially excluding cases from other features. Hence, solely minimizing the test suite is insufficient.
	
	In our work, we propose a scalable solution that also enhances coverage diversity, aiming to increase the likelihood of detecting unknown faults in larger test suites. We address this by ensuring the minimized test suite covers a more diverse set of features. The next section outlines the foundational problem formulation of our approach.

	\section{Problem Formulation}\label{model}
	We adopt the three criteria for the MCTSM problem described earlier: the size of the reduced test suite, statement coverage, and the number of faults detected. Recognizing the critical role fault detection plays in assessing the efficacy of a test suite \cite{rothermel1998empirical, jeffrey2007improving}, we treat the number of revealed faults as an \textit{absolute} criterion. In addition to minimizing the test suite size, we introduce another objective: minimizing pairwise coverage-based similarity. Our new formulation ensures that all known faults detectable by the original test suite continue to be detected by the minimized version without any omission. It offers a more straightforward solution to Nemo's aim \cite{lin2018nemo} of maximizing the number of \textit{distinct} known faults, eliminating the need for complex non-linear models. Moreover, it enhances coverage diversity, thereby increasing the potential to detect unknown faults. We employ two representations for the MCTSM problem: an integer linear program and a bipartite graph. Both representations serve as pivotal elements for our subsequent reinforcement learning approach.
	
	\subsection{ILP Formulation}\label{ilp_form}
	In our proposed Integer Linear Programming (ILP) formulation, represented by Equations \ref{ilp_4} through \ref{ilp_8}, we introduce additional considerations beyond statement coverage and fault detection. 
	
	{\small
		\begin{align}
		\textrm{Minimize  } &\sum_{1 \leq i \leq |T|} t_{i}  + \sum_{1 \leq i<j \leq |T|} c_{ij}\times y_{ij}\label{ilp_4}\\
		\textrm{Subject To:  } 
		&\sum_{1 \leq i \leq |T|} s_{ip}t_i \geq 1, 1 \leq p \leq |S| \label{ilp_5}\\
		&\sum_{1 \leq i \leq |T|} f_{ik}t_i \geq 1, 1 \leq k \leq |F|\label{ilp_6}\\
		&\left.\begin{aligned}
		y_{ij} & \geq x_i + x_j -1 \\
		y_{ij} &\leq x_i \\
		y_{ij} &\leq x_j\\
		\end{aligned}\right\} \forall i,j (1\leq i<j\leq |T|)\label{ilp_7}\\
		&t_{i}, y_{ij}\in \{0,1\}, 1\leq i<j\leq |T|\label{ilp_8}
		\end{align}
	}%
	
	Alongside the statement coverage matrix \( (s_{ip})_{1 \leq i \leq |T|, 1 \leq p \leq |S|} \) and the fault matrix \( (f_{ik})_{1 \leq i \leq |T|, 1 \leq k \leq |F|} \), we incorporate the similarity matrix \( (c_{ij})_{1 \leq i<j \leq |T|} \). This matrix quantifies the similarity between pairs of test cases regarding covered statements and faults. We introduce binary variables \(y_{ij}\) to represent the inclusion of the pair of test cases \(t_i\) and \(t_j\) in the reduced test suite. Specifically, \(y_{ij}\) equals to 1 when both \(t_i\) and \(t_j\) are included in the reduced suite; otherwise, it is 0. The objective function (equation \ref{ilp_4}) minimizes both the total number of tests ($\sum_{i=1}^{|T|} t_{i}$) and the total similarity score for the tests ($\sum_{1 \leq i<j \leq |T|} c_{ij}\times y_{ij}$) in the reduced suite. 
	
	Equation \ref{ilp_5} ensures coverage criteria. Specifically, it guarantees that each of the $|S|$ statements in the program is executed by at least one test case in the minimized suite. Similarly, equation \ref{ilp_6} confirms the fault detection criterion, stipulating that each of the $|F|$ known faults is detected by at least one test case in the reduced suite. Equation \ref{ilp_7} imposes linear constraints to ensure that \(y_{ij}\) equals to 1 only when both \(t_i\) and \(t_j\) are included in the reduced test suite. Lastly, equation \ref{ilp_8} defines the binary nature of the decision variables $t_i$ and  \(y_{ij}\). For variable $t_i$, a value of 1 indicates the inclusion of test case $t_i$ in the reduced suite, while 0 denotes its exclusion.
	
	\subsection{Bipartite Graph Formulation}\label{bi_form}
	Our key insight is that the MCTSM problem can be effectively represented as a bipartite graph, in addition to the above ILP formulation. Our rationale is that MCTSM can be reduced to a \textit{Set Covering} problem~\cite{hsu2009mints}, which in turn can be modelled as a bipartite graph~\cite{khalil2017learning}. 
	To our knowledge, we are the first to formulate the MCTSM problem as a bipartite graph.
	
	In a bipartite graph, nodes are separated into two distinct sets, and edges are only permitted between nodes from different sets. This clear separation ensures that no two nodes within the same set are directly connected. The Set Covering problem can then be defined within a bipartite graph as follows:
	
	\begin{definition}[Set Covering Problem]\label{scp}
		Given a bipartite graph \( \mathcal{G} = \{\mathcal{U}, \mathcal{V}, \mathcal{E}\} \), where:
		\begin{itemize}
			\item \( \mathcal{N} := \mathcal{U} \cup \mathcal{V} \) represents the combined node set,
			\item \( \mathcal{E} \) represents the edge set connecting nodes \( u \in \mathcal{U} \) and \( v \in \mathcal{V} \).
		\end{itemize}
		identify a minimized subset \( \mathcal{C} \subseteq \mathcal{U} \) such that every node in \( \mathcal{V} \) is connected to at least one node in \( \mathcal{C} \). Formally, $\forall v \in \mathcal{V} \Leftrightarrow \exists c \in \mathcal{C}$ and $(c, v) \in \mathcal{E}$, with the objective to minimize \( |\mathcal{C}| \).
	\end{definition}
	
	\begin{figure}[!t]
		\centering
		\includegraphics[width=2.2in]{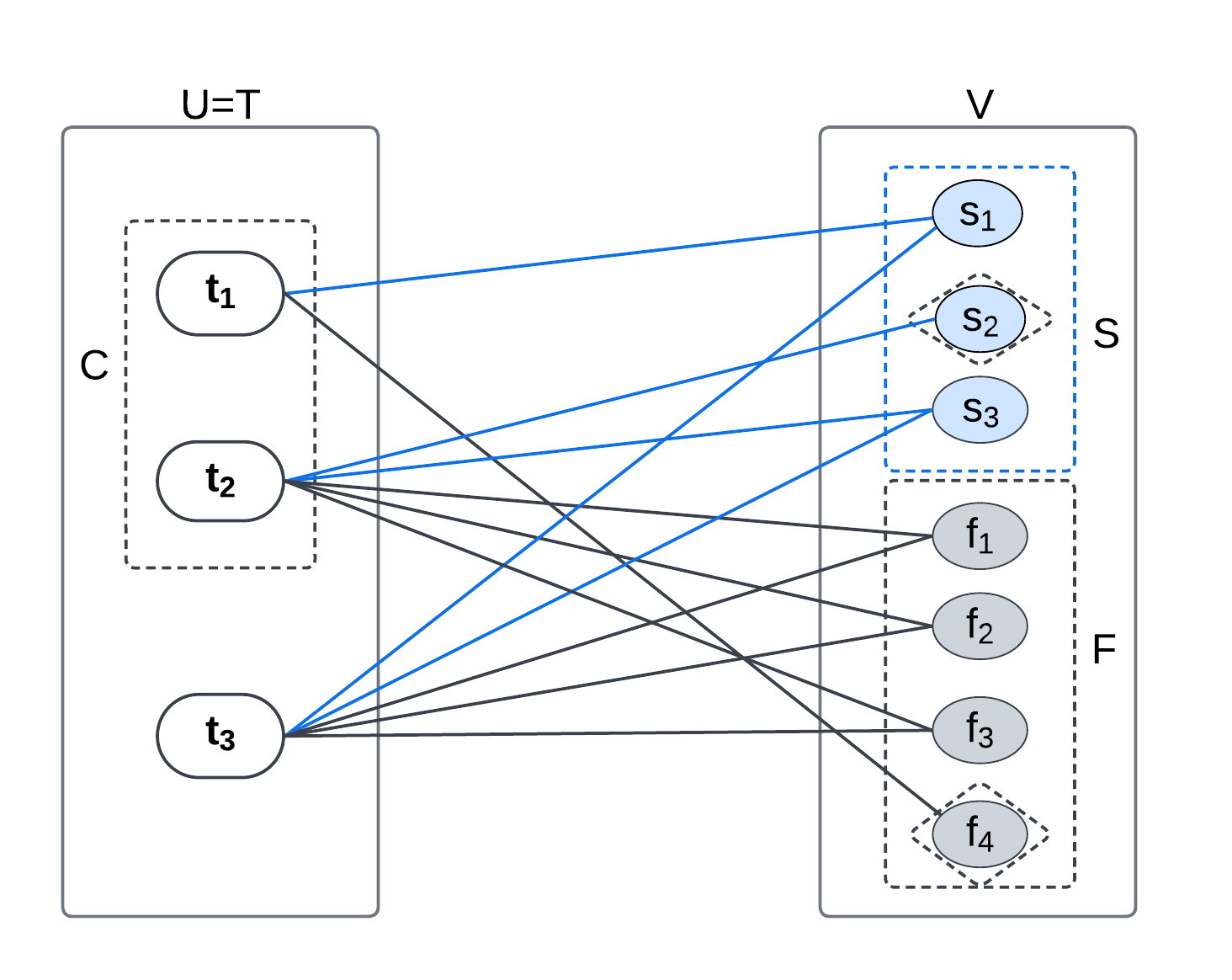}
		\caption{Bipartite Graph representation for Table \ref{tab:example}}
		\label{bipartite}
	\end{figure}
	
	We extend Definition \ref{scp} to the MCTSM problem where the node set $\mathcal{U}$ stands for all test cases, i.e., $\mathcal{U} := T$, while $\mathcal{V}$ includes all statements and faults, i.e., $\mathcal{V} := S \cup F$. For the edge set $\mathcal{E}$, each edge is either $(t, s) \in \mathcal{E}$ or $(t, f) \in \mathcal{E}$ where $t \in T, s \in S, f \in F$. Therefore, our MCTSM problem is equivalent to the bipartite graph problem to find the most diverse and smallest subset of $T$ that covers every node in $\mathcal{V} :=S \cup F$.
	
	As an example, a bipartite graph representation of the TSM instance from Table \ref{tab:example} is provided in Figure \ref{bipartite}. On the left side, enclosed within the large rectangle, lies the node set \( \mathcal{U} \), representing test cases \( t_1, t_2, \) and \( t_3 \) from the set \(T\). Meanwhile, on the right side within the separate rectangle is the node set \( \mathcal{V} \) which encompasses all statements and faults. The statement nodes are denoted in blue-dotted rectangle \(S\), whereas fault nodes are depicted in gray-dotted rectangle \(F\).
	
	The edges in this graph directly correlate with the information from Table \ref{tab:example}. As an illustration, an edge exists between \( t_1 \) and \( f_4 \), and notably, it's the sole edge linking \( f_4 \) to \( \mathcal{U} \). In Table \ref{tab:example}, this relationship corresponds to the single entry of 1 in the \( f_4 \) column against \( t_1 \). 
	
	Examining Figure \ref{bipartite}, two connections are exclusive: the edge \( (t_2, s_2) \) is the only link to \( s_2 \) and similarly, \( (t_1, f_4) \) is the unique connection to \( f_4 \). To ensure that all nodes in \( \mathcal{V} \) are covered, it's imperative that the subset \( \mathcal{C} \) of \( \mathcal{U} \) encompasses the nodes \( \{t_1, t_2\} \). Reviewing all the other nodes connected to either \( t_1 \) or \( t_2 \), it becomes apparent that they account for the entirety of \( \mathcal{V} \). Additionally, considering the similarity of test pairs, \(t_3\) is more similar to \(t_2\) than \(t_1\), considering the covered statements and faults. Consequently, the optimal solution should comprise \( \{t_1, t_2\} \). This solution aligns with the results obtained from our ILP formulation.

	\section{Approach}\label{approach}
\begin{figure}
		\includegraphics[width=0.7\textwidth]{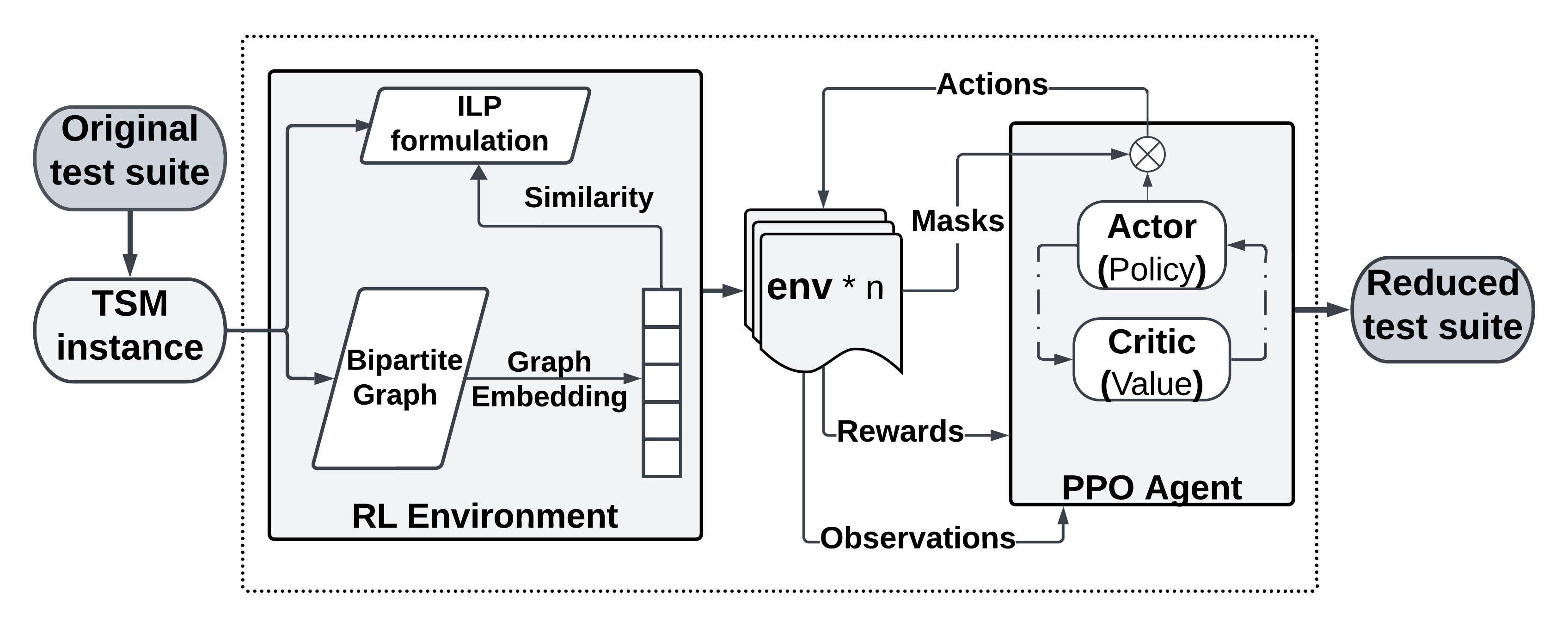}
		\caption{Overview of \toolname.}
		\label{overview}
	\end{figure}

	Reinforcement Learning (RL) has emerged as a promising framework for addressing combinatorial optimization problems \cite{mazyavkina2021reinforcement, nie2023reinforcement}, offering the ability to automatically learn effective heuristics without extensive domain-specific knowledge \cite{barrett2020exploratory}. Leveraging both ILP and bipartite graph formulations, we introduce \toolname, a novel approach focused on training an RL agent using a combination of a bipartite graph and ILP environment. This approach targets the practical MCTSM problem within the software engineering domain, aiming to efficiently generate reduced test suites. Figure \ref{overview} depicts an overview of \toolname. The essential components of the RL environment, including the ILP and bipartite graph formulations, have been covered in Section \ref{model}. We elaborate on the remaining modules concerning the design of the RL vectorized environment and the RL agent training procedure in the following subsections.
	
	\subsection{RL Environment}
	The bespoke RL environment for \toolname is developed based on the ILP and bipartite graph formulations (Section \ref{model}). As depicted in Figure \ref{overview}, this bipartite graph, along with its associated graph embeddings, constitutes the foundation of the RL environment. These bipartite graph embeddings are utilized to compute the similarity matrix essential for the ILP model, thereby providing guidance to the RL learning process. We then adopt a vectorized architecture by stacking multiple instances of this environment, a strategic decision aimed at significantly enhancing both time and sample efficiency during the RL training process. Below, we delve into the rationale of our design decisions concerning graph embedding, similarity measurement, and the vectorized environment.
	
	\noindent\textbf{Bipartite Graph Embedding.} To solve the MCTSM problem, the RL agent is trained to interact with the environment and make decisions regarding the inclusion of nodes from the set $\mathcal{U}$ (representing test cases) in the reduced test suite while ensuring that the original coverage of all nodes in $\mathcal{V}$ (representing statements and faults) is preserved. Therefore, the agent needs to extract the structural information about the bipartite graph including nodes, edges, and edge weights. However, the graph data is too large and complex for the agent to directly use for learning. To facilitate the learning process, we can transform graphs into compact embedding vectors, which capture the underlying topological characteristics in lower dimensions. Notably, widely-used graph embedding techniques such as \textit{node2vec}~\cite{grover2016node2vec} or \textit{graph2vec}~\cite{narayanan2017graph2vec} are typically designed for general graphs, overlooking the unique bipartite structures characterized by two distinct node types and inter-set edges~\cite{yang2022scalable, gao2018bine, cao2021bipartite}. In addition, existing RL-based approaches usually employ graph neural networks (GNN) to generate graph embeddings \cite{khalil2017learning, li2018combinatorial, barrett2020exploratory}. However, those approaches are not scalable to large graphs due to repeated iterations of GNNs.
	
	To address this challenge and effectively embed our bipartite graph environment, we employ the state-of-the-art \textit{bipartite network embedding} algorithm known as $\text{GEBE}^\text{p}$ (Poisson-instantiated \underline{GE}neric \underline{B}ipartite Network \underline{E}mbedding) \cite{yang2022scalable}. $\text{GEBE}^\text{p}$ attains the k-dimensional embedding vectors, denoted as $\mathbf{U}[u_i]$ for nodes within the set $\mathcal{U}$ and $\mathbf{V}[v_j]$ for nodes in $\mathcal{V}$, through performing singular value decomposition (SVD) \cite{musco2015randomized} over edge weight matrix $\mathbf{W}$. This algorithm is specifically designed to preserve not only the direct connections between nodes of different types but also the intricate multi-hop relationships, including \textit{multi-hop homogeneous similarity} (MHS) between nodes of the same type and \textit{multi-hop heterogeneous proximity} (MHP) between nodes of different types. Crucially, $\text{GEBE}^\text{p}$ boasts an impressive near-linear time complexity in relation to the size of the bipartite graph. This characteristic ensures its scalable and effective performance, even when dealing with large-scale bipartite graphs. Such scalability aligns perfectly with our research goals, allowing us to tackle complex MCTSM problems efficiently and effectively.
	
	\noindent\textbf{Pairwise similarity measurement.} Upon obtaining the bipartite graph embedding, we compute the necessary similarity matrix \( (c_{ij})_{1 \leq i<j \leq |T|} \) for the ILP objective function. For each test case \(t_i\), its corresponding embedding vector is denoted as $\mathbf{U}[t_i]$. We use cosine similarity to compute pairwise similarity as it is well-suited for high-dimensional data, such as our bipartite graph. Its bounded range between -1 and 1 aligns well with the normalized reward used in our RL training. Consequently, we calculate the cosine similarity score for each pair of test cases \((t_i, t_j)\) using their embedding vectors as $\frac{<\mathbf{U}[t_i], \mathbf{U}[t_j]>}{||\mathbf{U}[t_i]||*||\mathbf{U}[t_j]||}$ and take the absolute value to ensure the range is normalized to [0, 1]. This similarity score plays a pivotal role in determining the current value of the ILP objective function and serves as the basis for designing the reward function during RL training.
	
	\noindent\textbf{Vectorized Environment.} Moreover, to enhance time and sample efficiency, as well as optimize GPU/CPU resource utilization, we employ an efficient paradigm known as the \textit{vectorized environment} architecture \cite{towers_gymnasium_2023, stable-baselines3} that features a single learner to collect samples and learn from multiple independent environments, which are organized into a unified vectorized environment, either sequentially or in parallel. For each learning step, data such as observations and rewards collected from each environment are batched into vectors and passed to the agent for further training. This approach results in a speed-up by allowing multiple sub-environments to be sampled in parallel. As illustrated in Figure \ref{overview}, the ILP and bipartite graph formulations of the TSM instance, along with the extracted embedding vectors, are copied into $n$ independent environments for subsequent RL training.
	
	\subsection{RL Training}
	In the previous section, we introduced the environment in a static context before delving into its interaction with the RL agent. In this section, we will elaborate on the dynamic interaction between the environment and the RL agent, as well as our design decisions regarding observation representation, reward function, and action. The training algorithm is summarized in Algorithm \ref{ppo}.
	
	\noindent\textbf{Observation representation.}
	At each step of interaction between the environment and the agent, the current \textit{state} of the environment can be observed by the agent. An \textit{observation} is a partial description of a state that may omit certain details \cite{SpinningUp2018}. In our case, the graph embedding vectors generated by $\text{GEBE}^\text{p}$, which approximately capture the structure of the bipartite graph environment, serve as the observation. The embedding vectors $\mathbf{U}$ for nodes in $\mathcal{U}$ form a matrix with dimensions $|\mathcal{U}|\times k$, while $\mathbf{V}$ is a matrix with dimensions $|\mathcal{V}|\times k$. To represent the observation as a single real-valued vector, we opt for using the sum of all embedding vectors as the observation representation, which is $\sum_{u \in \mathcal{U}}\mathbf{U}[u] + \sum_{v \in \mathcal{V}}\mathbf{V}[v]$, where $\mathbf{U}[u]$ represents the embedding vector for a node $u \in \mathcal{U}$, $\mathbf{V}[v]$ denotes the embedding vector for a node $v \in \mathcal{V}$. As a result, the observation representation is a single vector in k-dimensional space, and this approach can be generalized to different-sized graphs \cite{khalil2017learning}. 
	
	When the agent takes an action, it leads to a corresponding change in the state of the environment. In the context of our MCTSM problem, an action involves selecting a node $u$ (i.e., a test case) from $\mathcal{U}$. Once $u$ is chosen, the corresponding state of the bipartite graph environment is updated as follows: node $u$ is marked as ``selected'' and its connected nodes in $\mathcal{V}$ (i.e., statements/faults) are marked as ``covered''. To effect this update, we modify the embedding vectors of the affected nodes to zeros values. Importantly, we opt for this approach rather than altering the graph's underlying topology and completely regenerating the embedding vectors from scratch. Because the impact on the embeddings of other nodes is minimal, it is unnecessary and time-consuming to regenerate the entire embedding matrix. Subsequently, the sum of all embedding vectors forms the observation for the current state and is passed to the agent.
	
	\noindent\textbf{Reward function.} The agent continues to take an \textit{action} for each step until the termination condition is met. In the context of our MCTSM problem, the termination condition is fulfilled when all nodes in $\mathcal{V}$ are covered. The primary objective of RL training is to minimize the selection of nodes in $\mathcal{U}$ and the total of similarity scores for selected nodes while preserving coverage for all nodes in $\mathcal{V}$. Therefore, the agent is trained to meet the termination condition as quickly as possible by taking fewer steps and minimizing the value of the ILP objective function simultaneously. 
	
	We design the immediate reward for each step as follows: \allowbreak $-\frac{selected\_U\_nodes}{|\mathcal{U}|} + \frac{distinct\_V\_nodes}{|\mathcal{V}|}$. Both $selected\_U\_nodes$ and $distinct\_V\_nodes$ are derived from the state of the environment. $-\frac{selected\_U\_nodes}{|\mathcal{U}|}$, as the first component, is negative. It represents the fraction of selected nodes in $\mathcal{U}$ up to the current step, divided by the total number of nodes in $\mathcal{U}$. This component encourages the agent to select fewer test cases since it yields a higher reward when the value of $selected\_U\_nodes$ is smaller. The second component, $\frac{distinct\_V\_nodes}{|\mathcal{V}|}$, is positive. It signifies the fraction of distinct nodes in $\mathcal{V}$ covered by the current action, divided by the total number of nodes in $\mathcal{V}$. A larger value is obtained when more distinct nodes are covered in the current step. The agent's objective is to maximize cumulative rewards, motivating it to learn that selecting fewer actions and choosing actions that cover more distinct nodes in each step are preferable. 
	
	Upon meeting the termination condition (i.e., when all nodes in $\mathcal{V}$ are covered), we refine the final reward by comparing the objective function of the best feasible solution encountered so far, denoted as $O(s^*)$, with the value of the ILP objective function of the current solution $O(s)$. The termination reward is defined as $\max \{O(s)-O(s^*),0\}$, providing a more informative signal to the RL agent, aiming at promoting the memorization of objective function improvements over time, contributing to more effective and exploratory learning \cite{barrett2020exploratory}. The sequence of states and actions taken is known as a \textit{trajectory}. If the termination condition is met, the trajectory is deemed feasible to our ILP model. However, in some cases where training is halted due to a timeout or a pre-defined step limit, the termination condition might not be reached, resulting in an infeasible trajectory. In such cases, we apply a penalty by subtracting 1 from the final reward.
	
	\noindent\textbf{Action. } In \toolname, the action space is discrete and its size corresponds to the size of the test suite. Originally, an action involves selecting a test case for inclusion in the reduced test suite. After formulating it into the bipartite graph, an \textit{action} in our problem is to select a node in $\mathcal{U}$ and add it to the partial solution, which accumulates selected nodes from previous steps. Therefore, the action space size is equal to $|\mathcal{U}|$. Each node in $\mathcal{U}$ can be selected at most once. After a node in $\mathcal{U}$ is selected or its connected nodes in $\mathcal{V}$ has already been covered by previous actions, the node becomes invalid for future actions. To prevent the agent from repeatedly sampling invalid actions, we employ the technique called \textit{invalid action masking} \cite{huang2020closer} to mask out the invalid nodes. For each step, the current \textit{mask} is generated based on the state of the environment and is passed to the agent along with the observation. The mask effectively restricts the agent's choice of actions to only valid nodes, excluding those that have been marked as invalid due to prior selections or coverage. The invalid action masking technique has proven to be a critical technique to accelerate the training process and outperform other agents that do not employ it \cite{shengyi2022the37implementation, huang2021gym}. Additionally, for problems with large discrete action spaces, this technique empowers the agent to learn more efficiently and scales well \cite{huang2020closer}. Given the large size of test suites in our MCTSM problem, invalid action masking proves to be highly beneficial for training and optimizing our RL agent's performance.
	
	\begin{algorithm}[t]
		\SetAlgoLined
		\small{envs = VecEnv(num\_envs = N)\;\label{alg2_1}
			agent = MaskableActorCriticMLP\;\label{alg2_2}
			obs= envs.reset(), dones = zeros(N)\;\label{alg2_3}
			\While{n < \text{total\_timesteps} // $(N \times n\_steps)$}{ \label{while_0}
				\Comment{\small{Rollout Phase: collect rollouts into buffer, buffer size $N \times n\_steps$}}
				\While{step < n\_steps}{\label{rollout_0}
					actions, other\_stuff = agent(obs, action\_masks)\;
					next\_obs, rewards, next\_dones, infos = envs.step(actions)\Comment*[r]{step in N envs}
					
					buffer.add(obs, actions, rewards, dones, action\_masks, other\_stuff)\Comment*[r]{store data}
					obs = next\_obs, dones=next\_dones\;
				}\label{rollout_1}
				\Comment{Learning Phase: update policy using the currently gathered rollout buffer to maximize the PPO-Clip objective.}\label{learn_0}
				agent.learn(buffer)\;\label{learn_1}
		}}
		\caption{PPO Training}
		\label{ppo}
	\end{algorithm}

	\noindent\textbf{Algorithm.} Our RL agent is trained using the \textit{Proximal Policy Optimization} (PPO) \cite{stable-baselines3, schulman2017proximal} algorithm, one of the state-of-the-art algorithms categorized under Actor-Critic policy gradient algorithms \cite{konda1999actor}. PPO trains a stochastic policy in an on-policy manner. In training the RL agent for our MCTSM problem, our goal is to obtain a solution as quickly as possible to meet the scalability requirement, prioritizing wall time efficiency. PPO offers several advantages that make it a suitable choice for \toolname. PPO is known to be more stable and sample efficient than the simple policy-gradient algorithms \cite{schulman2017proximal}. Moreover, it typically trains faster in terms of wall time than off-policy algorithms \cite{SpinningUp2018}. PPO supports discrete action space and requires minimal hyperparameter tuning. It can leverage multiprocessing through vectorized environments, further reducing the wall clock training time \cite{stable-baselines3}. 
	
	PPO as an Actor-Critic algorithm employs two networks: the actor and the critic. We use simple Multilayer Perceptron (MLP) to structure both the critic and actor networks in \toolname. The actor is responsible for generating valid actions based on the current invalid action mask, by sampling from the latest policy which is represented as a probability distribution of next-step actions. On the other hand, the critic evaluates the action produced by the actor by computing the value function, which corresponds to cumulative rewards. The policy is updated in a manner that maximizes the PPO-Clip objective \cite{SpinningUp2018}. 
	
	We outline the training procedure in Algorithm \ref{ppo}. Line \ref{alg2_1} initializes N bipartite graph environments using the \textit{gym} API \cite{towers_gymnasium_2023} and wrapped them into a vectorized environment. Line \ref{alg2_2} initializes the agent with two separate MLP networks, one for the actor and another for the critic, while enabling invalid action masking. Line \ref{alg2_3} initializes \texttt{observation} and \texttt{dones} vectors for subsequent training. The initial observation is computed as the sum of embedding vectors, following our designated observation representation. The algorithm comprises two phases: the rollout phase and the learning phase. In the rollout phase (Line \ref{rollout_0} - Line \ref{rollout_1}), for each iteration, the agent interacts with each environment for $n\_steps$ to collect a series of trajectories that are stored in a buffer. The size of the buffer is $N\times n\_steps$. In the learning phase (Line \ref{learn_0}-Line \ref{learn_1}), the collected data in buffer are used to update the policy using the PPO algorithm \cite{schulman2017proximal}. The agent is trained for \textit{total\_timesteps} // $(N \times n\_steps)$ iterations (Line \ref{while_0}). All feasible trajectories are saved during training, and the best one is selected as the final solution for our MCTSM problem.

	\subsection{Implementation}\label{implementation}
	\toolname is implemented in Python. The codebase including the ILP formulation, bipartite graph formulation, and embedding generation is publicly available \cite{repo}. We implemented the RL agent and environment based on the Maskable PPO algorithm of the \textit{stable-baseline3} \cite{stable-baselines3} framework. 
	
	\section{Evaluation}\label{evaluation}
	
	We assess \toolname to answer the following four research questions:
	
	\begin{itemize}
		\item \textbf{RQ1 (effectiveness):} How effective are the reduced test suites generated by \toolname in terms of test suite size, statement coverage, and fault detection capability?
		\item \textbf{RQ2 (efficiency):} How efficient is \toolname in generating reduced test suites in terms of runtime?
		\item \textbf{RQ3 (scalability):} To what extent is \toolname scalable for handling large test suites?
		
		\item \textbf{RQ4 (ablation study):} How does each component of \toolname contribute to its overall efficacy?
	\end{itemize}
	
	\subsection{Experimental Setup}\label{setup}
	
	\begin{table}
		\caption{Open-source subjects from Defects4J}
		\label{tab:real-sub}
		\small
		\begin{center}
			\setlength\tabcolsep{2pt} 
			\begin{tabular}{l|l|rrr|rrr} \hline
				{\bf Identifier} & {\bf Project name}& {\bf \#Tests} & {\bf \#Faults} & {\bf \#Stmt} & $|\mathbf{U}|$& $|\mathbf{V}|$ & $|\mathbf{E}|$ \\ 
				\hline\hline
				Cli&commons-cli&327&24&616&327&640&43,447\\
				JxPath & commons-jxpath&374&22&5,020&374&5,042&503,781 \\
				Jackson&jackson-core&581&16&7,973&829&7,989&324,239 \\
				Compress & commons-compress&627&42&1,532&627&1,574&75,278\\
				Jsoup&jsoup&688&83&5,473&688&5,556&667,618 \\
				Chart&jfreechart&1,324&17&5,184&1,324&5,201&349,218 \\
				Time&joda-time&3,837&17&4,545&3,837&4,562&3,885,762 \\
				Lang&commons-lang&4,087&25&13,964&4,087&13,989&326,637 \\
				Math&commons-math&4,397&76&1,940&4,397&2,016&414,701 \\
				Closure&closure-compiler&5,870&101&826&5,870&927&1,404,788 \\
				\hline
			\end{tabular}
		\end{center}
	\end{table}
	
	\noindent\textbf{Subject programs.} We conducted evaluations of \toolname using real-world subjects from Defects4J~\cite{just2014defects4j}, which provides a collection of reproducible bugs and the corresponding buggy version repositories for each subject. We selected the latest buggy version of each subject program from Defects4J v2.0.1 in our evaluation. Certain projects, such as Chart in Defects4J, rely on deprecated dependencies even when the latest versions are selected. For these subjects, we made manual adjustments to ensure that all subject programs are compatible with Java 8, Maven v3.6.3, JUnit v4.13.2, and Pitest \cite{coles2016pit} v1.15.8. Subsequently, for each project, we computed the corresponding \textit{TSM instance} as matrices between the original test suite and the criteria data (see example in \autoref{tab:example}) using the program's production code, test suite, and the collection of known faults from the Defects4J dataset. To clarify, the known faults do not correspond to the number of active bugs in the Defects4J dataset. Due to software evolution, some tests associated with known faults from earlier versions have been removed. The number of known faults in a TSM instance represents existing faults whose associated test cases are still present in the latest buggy version. We generated a fault matrix to indicate the relationship between each test and fault, determining whether test $t$ is able to detect a fault $f$. To ensure an adequate number of faults for formulating the ILP model, we opted to exclude subjects with fewer than 15 known faults. The statement coverage matrix is obtained by extracting tests and statements from the code coverage reports generated by executing the test suite. We utilized GZoltar \cite{gzoltar}, an open-source tool for automated testing and debugging, as recommended by the Defects4J authors. GZoltar provides a list of test cases, a list of statements from all classes under test, and a binary statement coverage matrix, where each row represents the coverage of a specific test case. The updated Defects4J projects, along with their associated TSM instance data (i.e., fault and statement coverage matrices) and the scripts for preparing these TSM instances, are publicly available in our code repository \cite{repo}.

	Table \ref{tab:real-sub} illustrates the 10 open-source subjects in our evaluation. The subjects are sorted based on their test suite size in column \#Tests, with five featuring smaller test suites (<1000 test cases) and the remaining five having larger test suites (>1000 test cases). Additionally, we provide the number of known faults and statements separately in columns \#Faults and \#Stmt. The column $|U|$ denotes the node set $\mathcal{U}$ in the bipartite graph formulation, aligned with the test suite size. The column $|V|$ indicates the node set $\mathcal{V}$, which is the sum of \#Faults and \#Stmt. The number of edges between the two node sets $\mathcal{U}$ and $\mathcal{V}$, denoted as $|E|$ (see \autoref{bipartite}), indicates the relationship between tests and criteria: statements and faults. If there is an edge between test $t$ and statement $s$, it means test $t$ covers statement $s$. Similarly, if an edge exists between $t$ and fault $f$, it indicates that $t$ triggers $f$. These edges are equivalent to the statement coverage and fault matrices as described in the TSM instance. Thus, $|E|$ can be utilized to capture the size of the TSM instance. In our evaluation, the test suite size ranges from 327 to 5,870, while the size of the TSM instances falls within the scale of $10^5$ to $10^6$. Note that the scale of the subjects in our evaluation is 10 to 100 times larger than those in existing MCTSM works \cite{xue2020multi, lin2018nemo}.

	\noindent\textbf{Baseline and Parameter Settings.} Exiting MCTSM papers  \cite{hsu2009mints, xue2020multi, lin2018nemo}  focus mainly on the ILP modelling and directly use commercial ILP solvers such as CPLEX \cite{cplex} to solve their ILP models. CPLEX can generally provide the true optimum of the ILP model for small-to-medium-sized test suites. Therefore, we provide a variant of \toolname, called \textsc{TripLEX}, that uses the state-of-the-art ILP solver CPLEX 22.1.1 \cite{cplex} to solve \toolname's ILP model (described in Section \ref{ilp_form}) directly. We configure CPLEX with a 2-hour time limit and a 2\% gap tolerance, instructing CPLEX to stop when either the 2-hour limit is reached or a solution within 2\% of the optimal is found. 
	
	Additionally, we compare \toolname with two recent test minimization techniques, one black-box tool: ATM \cite{pan2023atm} and one white-box tool: Nemo \cite{lin2018nemo}. ATM, as the state-of-the-art black-box test minimization tool, aims to minimize test similarity of the reduced test suite and leverages the Defects4J dataset for its evaluation. Unlike \toolname, ATM relies solely on test code without knowledge of production code. While \toolname uses a lightweight similarity measurement, ATM requires significant preparation time to create similarity measures based on test code pairs. However, since their replication package includes the similarity measures for subjects in Defects4J, we ensure a fair comparison by evaluating \toolname and ATM solely on the optimization process, excluding the preparation time. We collect the similarity scores for the same test suites in our subject projects and configure ATM with its optimal setup, which employs combined similarity using Genetic Algorithm (GA) as detailed in \cite{pan2023atm}. On the other hand, Nemo, as a white-box test minimization tool, formulates MCTSM problem as a single-objective integer program and focuses on ensuring a diverse set of faults are covered. Since our approach also formulates MCTSM problem into a single-objective integer program, we opt to compare with Nemo rather than other multi-objective Integer programming (MOIP) approaches \cite{xue2020multi}. Nemo incorporates test suite size, statement coverage, and fault detection ability and uses a nonlinear weight function to maximize the detection of distinct faults and linearizes their nonlinear formulation (NF\_NS) by integrating auxiliary decision variables into an ILP formulation called NF\_LS. NF\_LS enables the problem to be addressed by ILP solvers, improving time efficiency compared to NF\_NS.  We compare \toolname with Nemo's linearized formulation NF\_LS. In contrast with \toolname's ILP formulation, NF\_LS has a loose constraint for fault detection ability and does not consider test similarity.
	
	To evaluate the performance of \toolname, we recorded results after 10,000 RL training steps, ensuring that the agent had extensively engaged with the environment and collected a diverse set of trajectories for policy learning. To mitigate the impact of randomness, we executed each tool---\toolname, \textsc{TripLEX}, ATM, and Nemo---repeatedly over five trials for each of the 10 subjects listed in Table \ref{tab:real-sub}.
	
	All of our experiments are conducted on a 10-core Ubuntu 20.04 machine, equipped with two NVIDIA RTX A5000 GPUs and 64 GB memory. The hyperparameters for the RL agent are tuned based on the smallest subject Cli and hold constant for all other subjects for consistency. We fix the embedding dimensionality $k$ at 128.  The actor and critic MLP network layers are configured as $64\times64$ and $128\times128$, respectively. We select the learning rate of 0.0003 and minibatch size of 32. Both the observation and reward are normalized. We use 5 environments running in parallel as the vectorized environment. To ensure fairness, CPLEX also utilizes 5 threads in parallel. We set $n\_steps$ (see Algorithm \ref{ppo}) to 500, representing the number of steps the agent interacts with each environment. 
	
	\subsection{RQ1: Reduced Test Suite Effectiveness}\label{RQ1}
	
	To assess the effectiveness of the reduced test suites generated by \toolname, we measure the criteria: test suite size, detected faults, and statement coverage before and after minimization. 
	
\begin{figure*}
		\begin{minipage}[c]{0.45\textwidth}
			\captionof{table}{Reduced test suite size}
			\label{tab:size}
			\small
			\setlength\tabcolsep{1pt} 
			\begin{tabular}{llrrrr} \toprule
				{\bf Identifier} &{\bf (Original)}&{\bf \toolname}&{\bf \textsc{TripLEX}}&{\bf ATM}&{\bf Nemo}\\  
				\hline
				Cli &327  &61 & 60 & 61 & 57  \\
				JxPath &374  & 123  & 117  & 123 & 122  \\
				Jackson &581 &268 & 261 & 268 & 491 \\
				Compress &627  & 95  & 91  & 95 & 69  \\
				Jsoup &688 & 267  & 261  & 267 & 255 \\
				Chart&1,324  & 137  & 129  & 137 & 126 \\
				\hline
				Time &3,837 & 354  &N/A&N/A& 300  \\
				Lang &4,087 & 1,654  &N/A & N/A & 1,600 \\
				Math &4,397  & 160 & N/A & N/A & 80 \\
				Closure&5,870 & 149  &N/A & N/A & 73 \\
				\bottomrule
			\end{tabular}
		\end{minipage}
		\hfill
		\begin{minipage}[c]{0.45\textwidth}
			\centering
			\includegraphics[width=\textwidth]{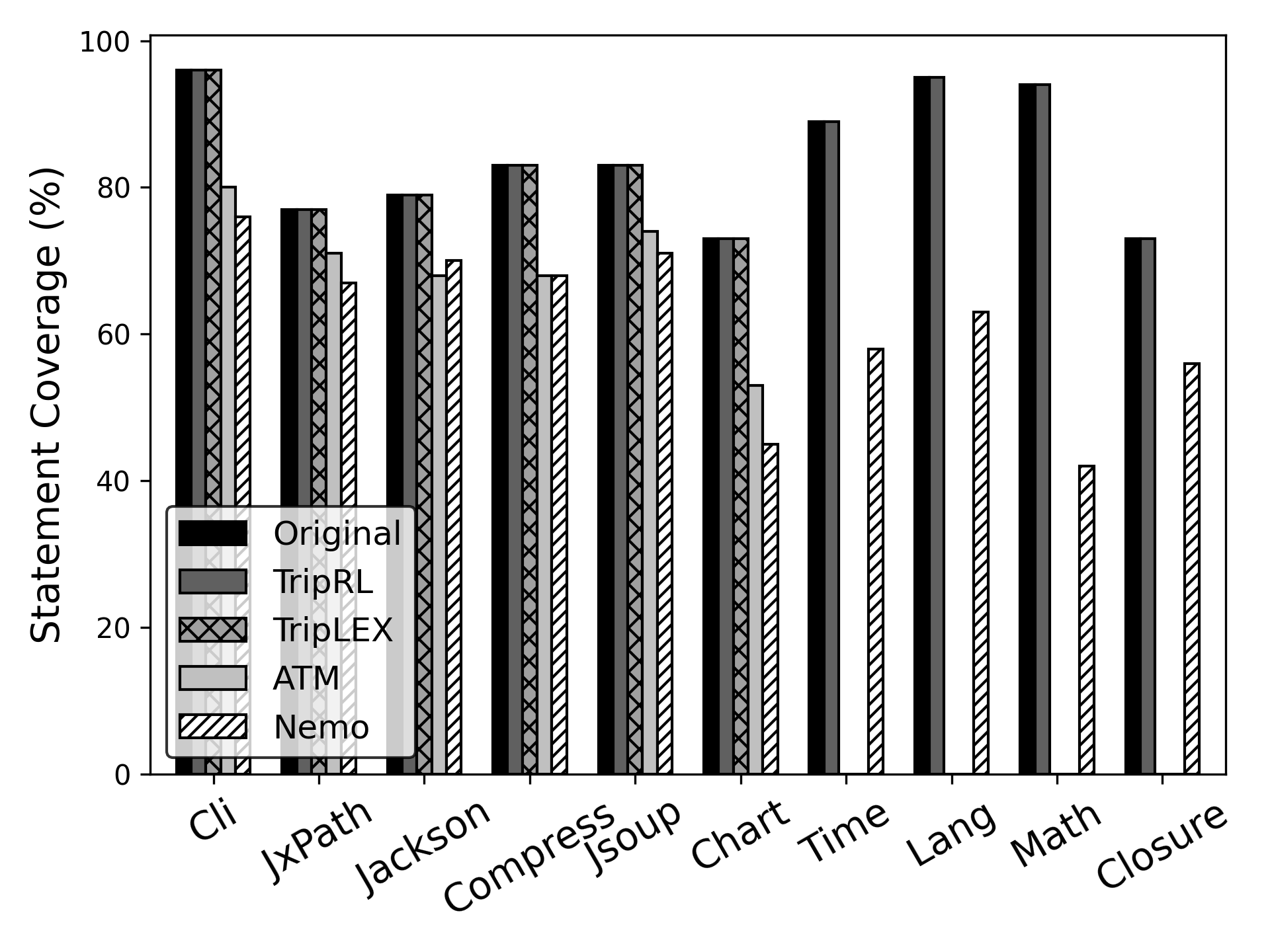}
			\setlength\abovecaptionskip{-10pt}
			\captionof{figure}{Statement coverage comparison}
			\label{scov}
		\end{minipage}
	\end{figure*}

	\noindent\textbf{Reduced test suite size.} We store the average reduced test suite size of the five trials for each subject in Table \ref{tab:size}. Column "Original" denotes the original test suite size before minimization. Then, we measure the size of reduced test suites by each technique. If a technique fails to find a solution within the 2-hour time limit or encounters an out-of-memory issue, we denote the result as N/A. All four techniques can reduce tests for small subjects with up to 1.3K test cases (i.e., Cli to Chart). However, \textsc{TripLEX} failed to reduce test suites for larger subjects due to either timeout or out-of-memory issues. In contrast, \toolname consistently generates reduced test suites across all 10 subjects, including those with larger test suites. When comparing the reduced test suite sizes between \toolname and \textsc{TripLEX}, which both follow our ILP formulation (described in Section \ref{ilp_form}), the reduced test suites generated by \toolname are slightly larger than those generated by \textsc{TripLEX}. The ratio between the size of \toolname and \textsc{TripLEX} ranges from 1.017 to 1.062 for the small subjects Cli to Chart. These results indicate that \toolname can generate near-optimal solutions competitive with the commercial ILP solver CPLEX. ATM, as a black-box technique, requires the user to provide a budget for the reduced test suite size. For a fair comparison, we calculate the budget as the percentage of \toolname's size over the original test suite size, ensuring that the size of the reduced test suite is consistent between TripRL and ATM. Similar to \textsc{TripLEX}, ATM fails to reduce test suites for the larger test suites due to timeout issues. In contrast, Nemo reduced test suites for all 10 subjects. The reduced test suites generated by Nemo follow the NF\_LS ILP formulation as discussed in Section \ref{setup}.
	
	\noindent\textbf{Statement Coverage.} Figure \ref{scov} illustrates bar plots for the statement coverage of the original and reduced test suites generated by each technique. The bars for both \toolname and \textsc{TripLEX} have the same height as the original in Figure \ref{scov}. This shows that both \toolname and \textsc{TripLEX} (for smaller test suites) maintain the same statement coverage as the original, aligning with our ILP formulation constraint (\ref{ilp_5}). In contrast, ATM and Nemo yield a significant loss in statement coverage, as illustrated in Figure \ref{scov} with shorter bars than the original. ATM's statement coverage loss ranges from 6\% for JxPath to 20\% for Chart, while Nemo performs worse than ATM, with statement coverage losses ranging from 9\% for Jackson to 52\% for Math.
	
\begin{figure*}
		\begin{minipage}[c]{0.45\textwidth}
			\centering
			\includegraphics[width=\textwidth]{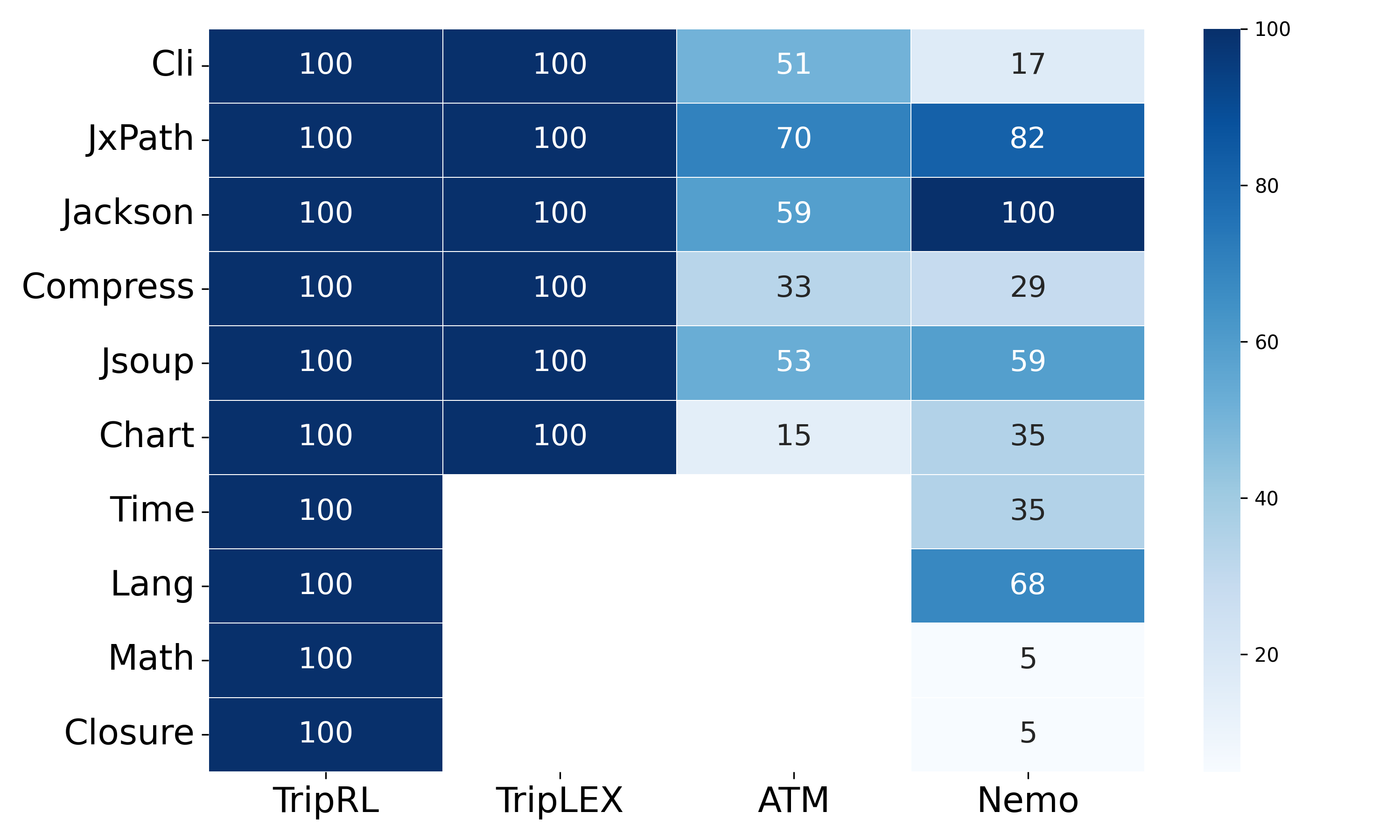}
			\setlength\abovecaptionskip{-5pt}
			\captionof{figure}{Fault detection rate comparison}
			\label{fdr}
		\end{minipage}
		\hfill
		\begin{minipage}[c]{0.5\textwidth}
			\includegraphics[width=\textwidth]{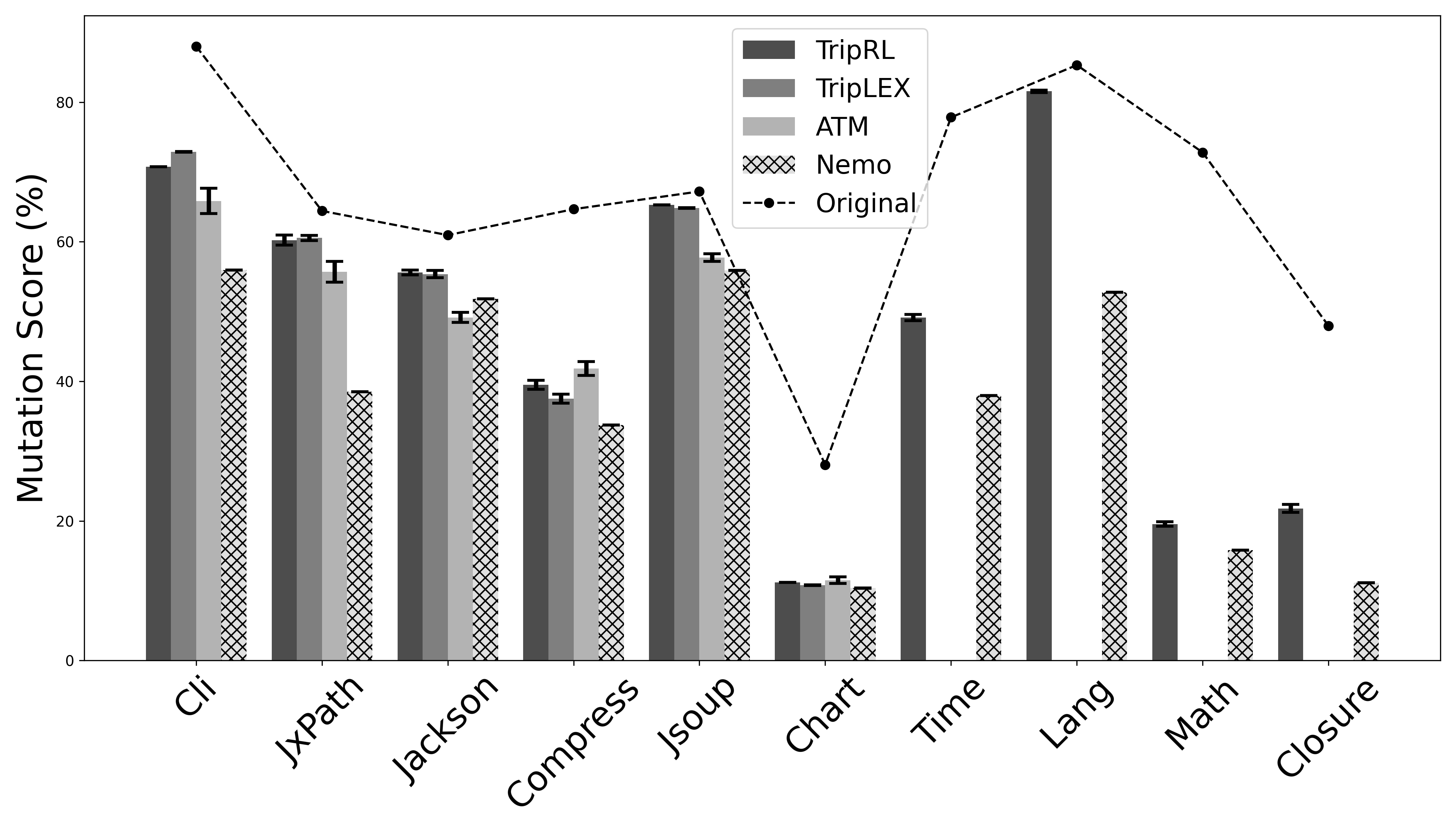}
			\captionof{figure}{Bar plot with error bars of mutation scores among different techniques}
			\label{mutation}
		\end{minipage}
	\end{figure*}

	\noindent\textbf{Fault Detection Ability.} We use the Fault Detection Rate (FDR) capturing the percentage of detected faults over the total known faults, to represent the fault detection ability of the reduced test suite. The FDR for the four techniques is illustrated as a heatmap in Figure \ref{fdr}, with blank spaces left for N/A values. Both \toolname and \textsc{TripLEX} (for the smaller test suites) achieve a 100\% fault detection rate, aligning with the ILP formulation's constraint (\ref{ilp_6}). In contrast, ATM and Nemo suffer from fault detection loss. 
	
	%
	
	However, the known faults are limited to only 423 in total in our evaluation. Achieving a 100\% fault detection rate for known faults cannot guarantee strong fault detection ability for future unknown faults. Therefore, we extend beyond the known faults to evaluate the test effectiveness of the reduced test suite in identifying future faults. We leverage the widely used mutation testing approach to simulate unknown faults and measure mutation scores \cite{just2014mutants} for the reduced test suites across different minimization techniques. We utilize the mutation testing tool PIT \cite{coles2016pit} to generate mutants, as it can generate the same set of mutants consistently, provided the production code remains unchanged. This ensures a fair comparison across different minimization techniques.
	
	We first obtain the mutation score of the original test suite before minimization for each subject, listed under Row "Original" in Table \ref{tab:FDA}. For the four minimization techniques, we calculate the average mutation score of the five trials for each subject. The average mutation scores for each technique are listed under Rows \toolname to Nemo in Table \ref{tab:FDA}. We compare \toolname with the other three techniques (\textsc{TripLEX}, ATM, Nemo) separately and calculate the percentage gain using the formula $(\text{\toolname's score} - \text{other tool's score})/\text{other tool's score} \times 100$. Note that the average values under "Ave" Column marked with * indicate that the value is calculated for the six subjects workable for the corresponding technique, not all 10 subjects. In addition to the average mutation scores for five trials in Table \ref{tab:FDA}, Figure \ref{mutation} is a bar plot with error bars representing the mutation scores. The height of each bar indicates the average mutation score (mean value), the same as the values in Table \ref{tab:FDA}, while the error bars indicate the standard deviation (i.e., the variability) across the five trials.
	
	Comparing \toolname to \textsc{TripLEX}, \textsc{TripLEX} achieves a slightly higher mutation score than \toolname for the smallest two subjects, Cli and JxPath. However, its performance drops with larger subjects. This result also demonstrates that our RL approach is more powerful than CPLEX when solving our ILP formulation for larger subjects. On average across the six subjects workable for \textsc{TripLEX}, \toolname still has a 1.11 percentage gain regarding mutation score. Compared to ATM, on average, \toolname has a 5.58 percentage increase in mutation score. When considering each subject individually, ATM obtains a higher mutation score than \toolname for the subjects Compress and Chart however it performs worse for the remaining subjects. It is worth noting that the error bars for ATM in Figure \ref{mutation} are generally larger compared to other tools. This indicates that the mutation scores for ATM have higher variability across different runs, suggesting that ATM is more affected by randomness due to its lack of knowledge about the source code. Nemo is the most stable considering the error bars; however, Nemo consistently has much lower mutation scores than \toolname. \toolname outperforms Nemo with a 33.5 percentage gain. 
	
	Overall, \toolname outperforms the other techniques in test suite reduction and effectiveness. 
	
	\begin{table}
		\captionof{table}{Mutation scores (\%) of the original and reduced test suites}
		\label{tab:FDA}
		\footnotesize
		\begin{center}
			\setlength\tabcolsep{2pt} 
			\begin{tabular}{ll|cccccccccc|l} \toprule
				\multicolumn{2}{c|}{\diagbox{Techniques}{subject}}&Cli &JxPath & Jackson &Compress & Jsoup & Chart & Time & Lang & Math &Closure &Ave.\\
				\midrule
				\multirow{5}{*}{\rotatebox{90}{mutation score }}&(Original)  &88.01 &64.42 &60.95 & 64.67	&67.21 &28.05 &77.83 & 85.32 & 72.81 & 47.96 &65.72 \\
				\cmidrule{2-13}
				&\toolname   &70.75 & 60.24 &55.61 & 39.50 &65.28 & 11.19 &49.14 &81.56 &19.56 & 21.79 &47.46\\
				&\textsc{TripLEX} &72.88 &60.55 &55.37 &37.51 &64.83 &10.79 &N/A&N/A&N/A& N/A&50.32* \\
				&ATM&65.85 &55.69 &49.16&41.84	&57.73	&11.50 &N/A&N/A&N/A& N/A &46.96*\\
				&Nemo&55.93 & 38.51 &51.83 &33.73 &55.89 &10.37 &37.96 & 52.77 & 15.82 & 11.16&36.40\\
				\midrule
				\multirow{3}{*}{\rotatebox{90}{ \% Gain}}&\toolname over \textsc{TripLEX} &-2.93 & -0.51 &0.43 & 5.29 & 0.69 &3.71&N/A&N/A&N/A& N/A &1.11*\\
				&\toolname over ATM &7.44& 8.16 & 13.11 & -5.58& 13.09  & -2.71 &N/A&N/A&N/A& N/A & 5.58*\\
				&\toolname over Nemo &26.50 & 56.40 &  7.29 & 17.10 &16.80 &7.91 &29.47 &54.55&23.66&95.32 &33.50\\
				\bottomrule
			\end{tabular}
		\end{center}
	\end{table}
	
	\subsection{RQ2: Time efficiency}\label{time}
	
	We compared the time efficiency among the techniques, as summarized in Table \ref{tab:time}. Nemo is the fastest approach, providing a minimized test suite in less than 1 second for all the subjects. For the remaining three techniques, \textsc{TripLEX} proves to be the most efficient solution for generating reduced test suites for small subjects. In cases with fewer than 1,000 test cases, \textsc{TripLEX} generates solutions in less than 144 seconds. The only exception is the subject JxPath, where ATM is the fastest, but \textsc{TripLEX} is still competitive with ATM for JxPath. ATM is worse than either \toolname or \textsc{TripLEX} for all the other subjects. 
	
	For \textsc{TripLEX}, the execution time of the ILP solver CPLEX \cite{cplex} grows exponentially with the size of test suite and criteria data matrices. For ATM, the execution time of their search algorithm increases quadratically with the number of test cases \cite{pan2023atm}. As the test suite size increases beyond 1,000 tests, the time required for both \textsc{TripLEX} and ATM increases substantially. For instance, for Chart with 1,324 tests  \toolname completes in 568.22 seconds while \textsc{TripLEX} requires 1679.52 seconds to run, making it approximately three times slower than \toolname. ATM requires 771.69 seconds for the subject Chart which is around 36\% slower than \toolname. For the other four larger subjects with test suite sizes ranging from 3,837 to 5,870 test cases, both ATM and \textsc{TripLEX} are unable to generate solutions within the two-hour time limit or are terminated due to out-of-memory errors. \textsc{TripLEX} fails to generate output for subjects Lang and Math within the two-hour time limit. Even after removing the time limit, it still cannot provide a solution after running overnight (exceeding 15 hours). Additionally, for subjects Time and Closure, \textsc{TripLEX} fails to execute due to an out-of-memory error while processing the ILP models. 
	
	In contrast, \toolname provides solutions consistently for all subjects in at most less than 47 minutes. It is worth noting that the time for \toolname includes the embedding and similarity generation, creation of the vectorized environment and the duration for 10,000 steps of training. 
	
	\begin{table}
		\caption{Time efficiency (seconds)}
		\label{tab:time}
		\small
		\begin{center}
			\begin{tabular}{lrrr|rrrr} \hline
				{\bf Identifier} & {\bf \#Tests} & {\bf \#Stmt} & $|\mathbf{E}|$ &{\bf \toolname} &{\bf \textsc{TripLEX}} &{\bf ATM} & {\bf Nemo}\\ 
				\hline\hline
				Cli&327&616&43,447 &119.46 &21.71  &47.97& 1\\
				JxPath &374&5,020&503,781 &598.36 &141.14  & 85.69 & 1\\
				Jackson& 581 &7,973&324,239 &489.50 &128.87 & 254.14 &1\\
				Compress &627&1,532&75,278 &169.28 &143.17 &240.51 & 1\\
				Jsoup&688&5,473&667,618 &742.39 &122.47 &332.92 &1 \\
				Chart&1,324&5,184&349,218 &570.90 &1,679.52 &771.69 & 1\\
				\hline
				Time&3,837&4,545&3,885,762 &2,802.92 & N/A& N/A & 1 \\
				Lang&4,087&13,964&326,637 &1,291.36 & N/A& N/A  & 1\\
				Math&4,397&1,940&414,701 &876.60& N/A& N/A  & 1\\
				Closure&5,870&826&1,404,788 &2,136.77& N/A& N/A  & 1\\
				\hline
			\end{tabular}
		\end{center}
	\end{table}

	\subsection{RQ3: Scalability}
	The relationship between \toolname's runtime and test suite size, as observed in Table \ref{tab:time}, does indeed exhibit a notable positive correlation. However, unlike black-box tools such as ATM, test suite size alone does not fully determine the runtime for \toolname. For instance, although Closure boasts the largest test suite size of 5,870 tests, it does not necessitate the longest running period. Conversely, Time, despite having a smaller test suite of 3,837 tests compared to Closure, requires the most substantial amount of execution time. This discrepancy is likely due to Time's larger number of edges, as indicated in Column $|E|$, which denotes the size of the TSM instance. Additionally, Lang has fewer tests and edges compared to Math, yet its runtime is significantly higher. This may be attributed to Lang having a considerably larger number of statements from all classes under test, as indicated in Column \#Stmt in Table \ref{tab:time}. The number of statements can be used as a metric for the size of project's production code.
	
	We hypothesize that \toolname's runtime is influenced by the size of the test suite (\#Tests in Table \ref{tab:time}), the size of the TSM instance ($|E|$ in Table \ref{tab:time}), as well as the size of production code (\#Stmt in Table \ref{tab:time}). To investigate this hypothesis, we utilize a multiple linear regression model \cite{stojiljkovic2021linear}, with \toolname's runtime as the dependent variable and the number of tests, statements, and edges as the independent variables. With five trials conducted for each of the ten subjects in Table \ref{tab:real-sub}, we gathered a dataset of 50 samples to construct the regression model. The linear model is expressed as: $y = -8.77 + 15.59 x_1+ 39.33 x_2 + 53.73 x_3$, where $y$ signifies running time in seconds, $x_1$ represents the number of Tests scaled by a factor of $10^2$, $x_2$ denotes the number of statements scaled by a factor of $10^3$, $x_3$ denotes the number of edges scaled by a factor of $10^5$. The model exhibits an impressive R-squared value of 99\%, indicating a high degree of accuracy in capturing data variability. Thus, we can confidently conclude that \toolname's runtime shares a linear relationship with the three features: the number of tests, statements, and edges.
	
\begin{figure*}
	\caption{Pairwise relationships of the three features: \#Tests, \#Statements and \#Edges against \toolname's runtime }
	\label{scalability}
	\begin{minipage}[c]{0.3\textwidth}
	\includegraphics[width=\textwidth]{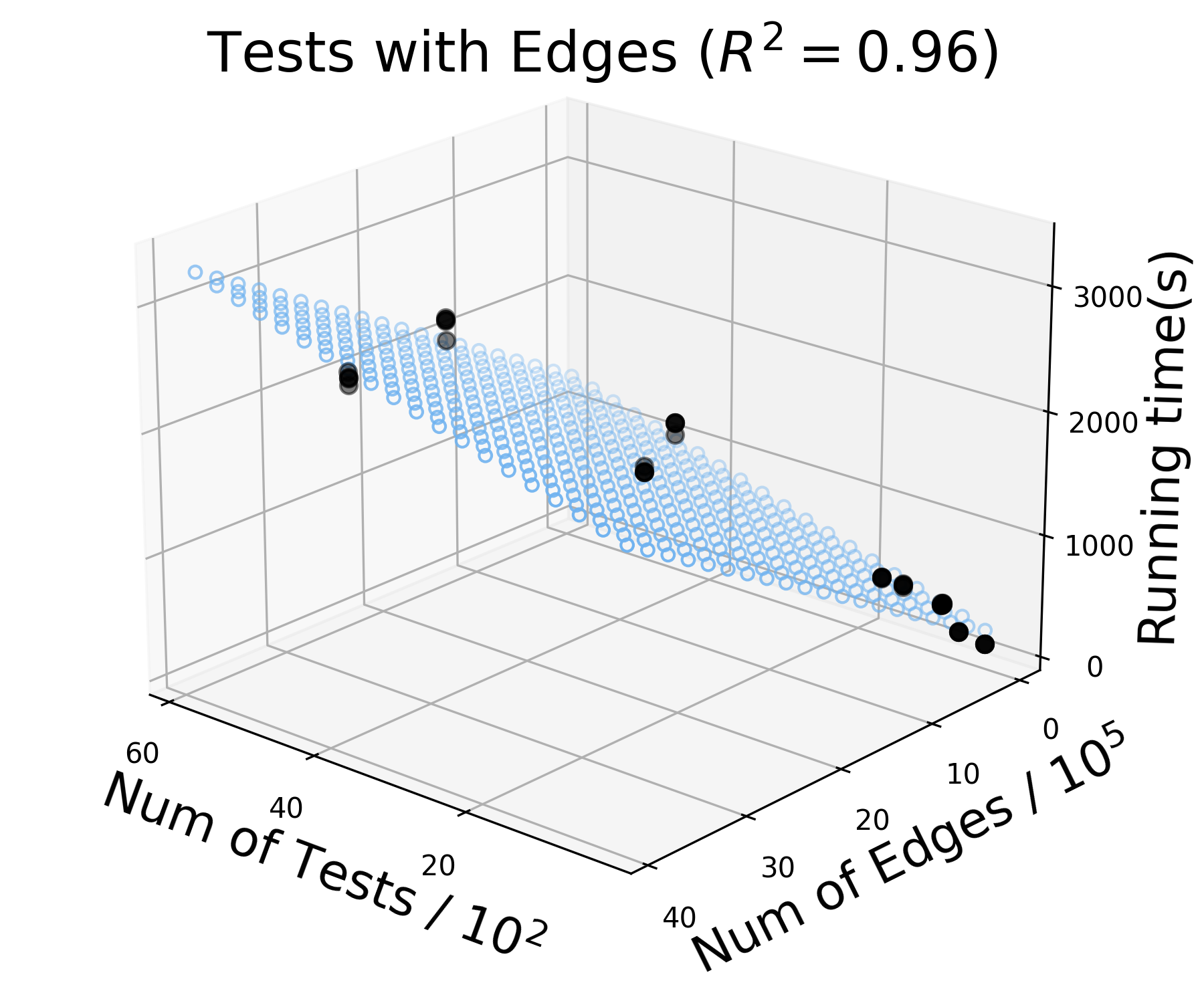}
	\end{minipage}
	\hfill
	\begin{minipage}[c]{0.3\textwidth}
			\includegraphics[width=\textwidth]{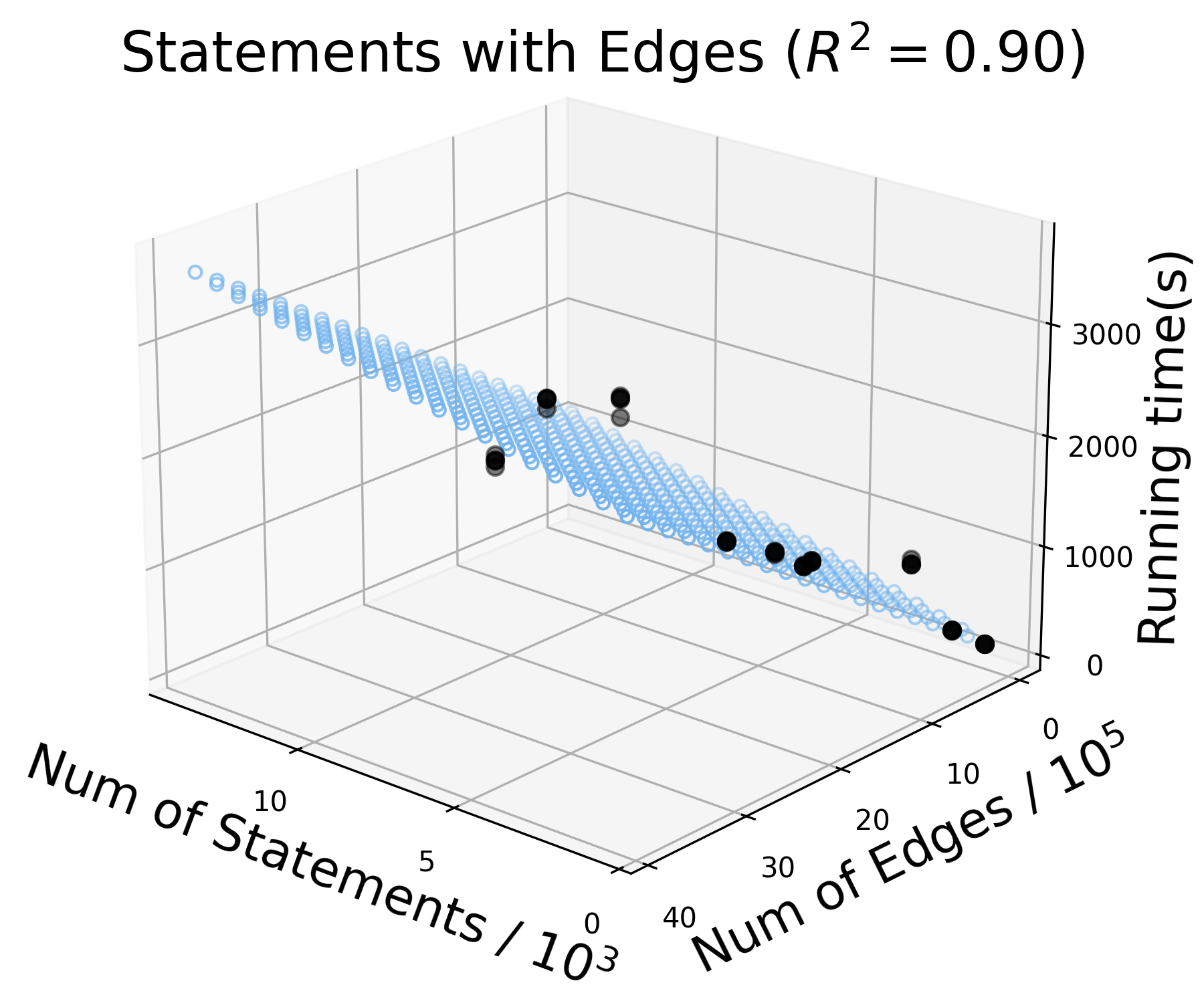}
	\end{minipage}
	\hfill
	\begin{minipage}[c]{0.3\textwidth}
		\includegraphics[width=\textwidth]{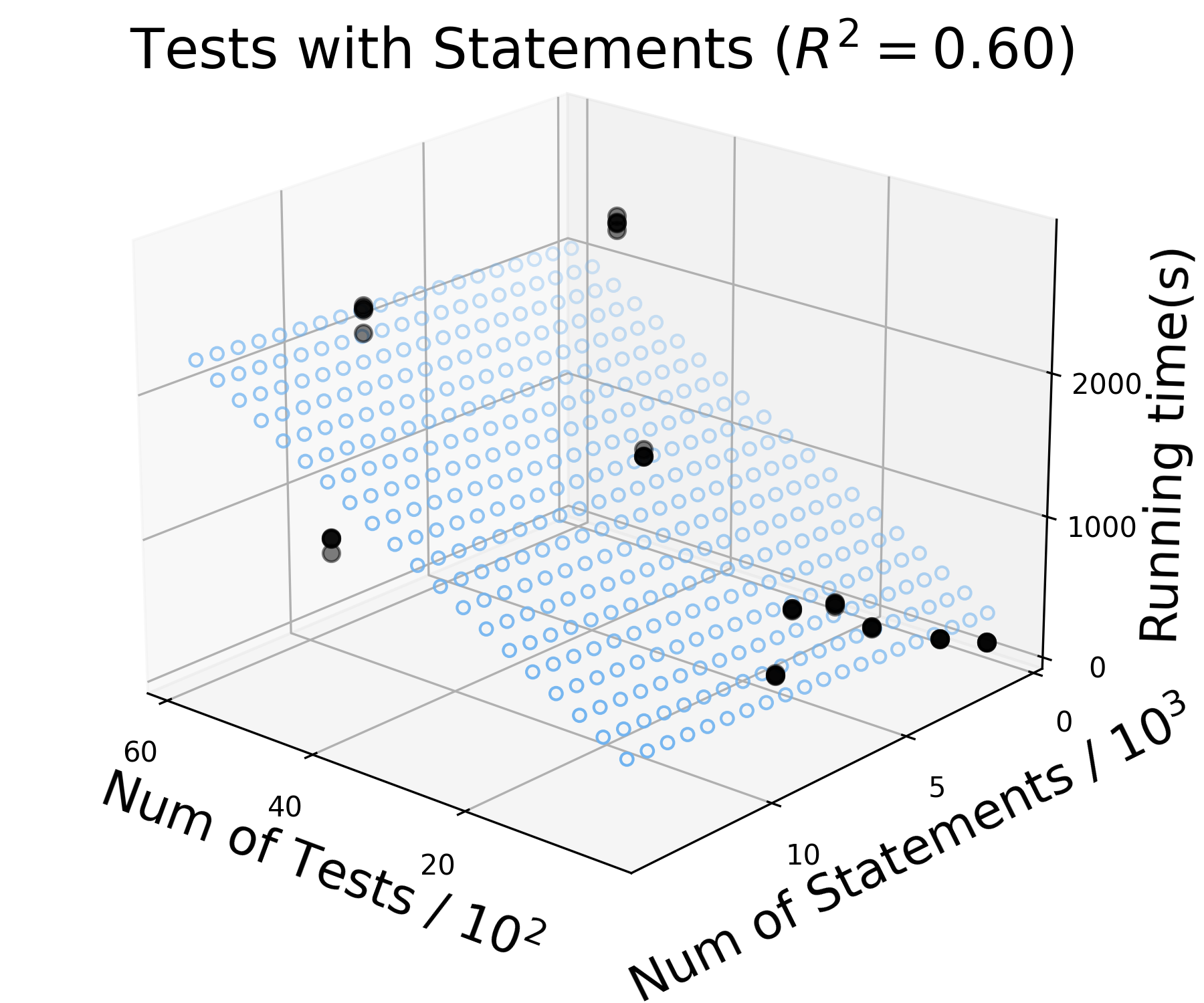}
	\end{minipage}
    \end{figure*}

	Additionally, to further evaluate the impact of the three features on runtime, we selected pairs of these features and fit a linear model for each pair. Figure \ref{scalability} provides a visualization of three feature pairs (Tests with Edges, Statements with Edges, and Tests with Statements) and reports their respective R-squared values. By excluding features such as Tests or Statements, the R-squared value dropped slightly, but the model still achieved an R-squared over 90\%. However, removing the edges resulted in a significant drop in the R-squared value from 99\% to 60\%. This indicates that the size of the TSM instance has the strongest impact on overall runtime, while the size of the test suite and production code have comparatively lower impacts on runtime.
	
	By leveraging the multiple linear regression model, we can predict the runtime for larger-scale software systems. To assess \toolname's scalability using this model, in considering a scenario where the test suite size of $10^4$, production code size as $10^5$ and the TSM instance size of $10^7$ represents an increase of one order of magnitude compared to the subjects in our current evaluation, \toolname is capable of reducing the test suite in approximately 3 hours, demonstrating its scalability for larger software projects.
	
	\subsection{RQ4: Ablation study}
    We evaluated \toolname against its variant, \textsc{TripLEX}, and demonstrated that our RL-based approach outperforms the ILP-based approach using CPLEX in terms of both time efficiency and scalability, as shown in the previous research questions. To further assess the contributions of specific components of \toolname, including the bipartite graph embedding and similarity measure, we conducted a detailed ablation study.
	
	For this study, we developed two variants of \toolname: `w/o bipartite' and `w/o similarity'. The `w/o bipartite' variant replaces the bipartite graph embedding with Node2Vec \cite{grover2016node2vec}, a widely used homogeneous graph embedding method. The `w/o similarity' variant removes the pairwise similarity measurement in the RL training process. Instead of using the cosine similarity measure (ranging from [0, 1]) employed by \toolname, The `w/o similarity' variant assigns a constant similarity value of 0.5 for all pairs. These variants allow us to isolate and evaluate the impact of each component on the overall performance of \toolname. Because the objective function is still the same to maintain the same statement coverage and fault detection ability, the results of reduced test suite size, statement coverage and fault detection rate remain the same as \toolname for the two variants. We report the runtime and mutation scores in Table \ref{tab:ablation}. 
	
	For the `w/o bipartite' variant, the runtime is significantly higher than \toolname and it failed to produce results for the larger subjects: Time, Lang, Math and Closure due to Timeout or Out-of-memory issues. This demonstrates that the bipartite graph embedding method, $\text{GEBE}^\text{p}$ \cite{yang2022scalable}, used in \toolname, is a more efficient and scalable embedding method compared to Node2Vec. In terms of mutation scores of the reduced test suite, \toolname achieves either higher or equal mutation scores for subjects Cli, JxPath and Jackson, Jsoup and Chart compared to the `w/o bipartite', with the only exception being the subject Compress. Therefore, we conclude that the bipartite graph embedding component in \toolname contributes significantly to time efficiency and scalability, and also enhances the potential fault detection capability for most of the subjects.
	
	For the `w/o similarity' variant, it efficiently generates reduced test suites for all subjects except Time. The subject Time failed due to an out-of-memory issue. The reason is that, with constant similarity scores assigned to all test pairs, the RL agent struggles to differentiate between test cases, making it difficult to prioritize or make meaningful selections. This leads to the agent exhausting its resources while attempting to select from the test cases. Regarding runtime, for smaller subjects from Cli to Chart, the difference between \toolname and the `w/o similarity' variant is marginal. For larger subjects, the `w/o similarity' variant is generally faster than \toolname; however, it faces an out-of-memory issue for the subject Time. In terms of mutation scores, \toolname achieves higher or equal mutation scores for all subjects compared to the `w/o similarity' variant. This indicates that \toolname has a higher potential to detect unknown faults in general. Thus, we can conclude that the pairwise similarity measurement component contributes significantly to fault detection capability, aids the RL agent in decision-making, and does so without incurring significantly longer runtime or consuming excessive memory resources.
	
    \begin{table}
		\caption{Ablation Study}
		\label{tab:ablation}
		\small
		\begin{center}
			\begin{tabular}{@{}lcccccc@{}} \toprule
				\multirow{2}{*}{\bf Identifier}&\multicolumn{3}{c}{\bf Time efficiency (seconds)}
				&\multicolumn{3}{c}{\bf Mutation scores (\%)}\\ \cmidrule(l){2-4}\cmidrule(l){5-7}
				& \toolname & w/o bipartite & w/o similarity&  \toolname & w/o bipartite & w/o similarity\\ \midrule
				Cli&119 &186& 117& 71&71&71\\
				JxPath &598&2,538 & 612&60 & 58& 58\\
				Jackson&490 & 1,560& 449&56&55&55\\
				Compress  &169 & 332&163 & 40&43&40\\
				Jsoup &742 &3,224& 741& 65&65&65\\
				Chart&571 &1,635& 553 &11 &11&11\\
				\hline
				Time&2,803 &N/A & N/A & 49&N/A&N/A \\
				Lang&1,291 & N/A& 1,098 & 82&N/A&82\\
				Math &877& N/A&  621& 20&N/A&19\\
				Closure &2,137& N/A& 1,704 & 22 &N/A&21\\
				\hline
			\end{tabular}
		\end{center}
    \end{table}
	
	\section{Discussion}
	
	\header{Effectiveness.} As our results show, \toolname demonstrates superior effectiveness compared to other techniques by consistently generating reduced test suites across all 10 subjects, including those with larger test suites, where \textsc{TripLEX} and ATM fail due to timeout or memory issues. It maintains the same statement coverage as the original test suites, unlike ATM and Nemo, which suffer significant coverage losses. Additionally, \toolname achieves a 100\% fault detection rate for known faults. 
	Nemo performs better than ATM in terms of fault detection rate of known faults: while ATM's FDR is at most 56\%, Nemo achieves a 100\% FDR for the subject Jackson and 82\% for JxPath. This aligns with Nemo's design purpose to maximize distinct fault detection.  However, because Nemo still treats fault detection as a relative criterion (see Section \ref{challenges}) instead of an absolute criterion, it cannot guarantee 100\% FDR for known faults across all subjects. In our ILP formulation, we enforce a constraint (Equation \ref{ilp_6}) that ensures all known faults are covered at least once by the reduced test suite. 
	
	When comparing the error bars in mutation scores shown in Figure \ref{mutation}, the order from largest to smallest is $\text{ATM} > \toolname \approx \textsc{TripLEX} > \text{Nemo}$. The error bars indicate variability among trials.  ATM exhibits less consistency and more fluctuation than the other three techniques. In contrast, ILP-based approaches including \toolname, \textsc{TripLEX}, and Nemo generally exhibit lower fluctuations. Nemo scores the lowest mutation scores while \toolname exhibits higher mutation scores across all subjects, indicating better potential for detecting unknown faults.
	
	\header{Efficiency and Scalability.} Nemo is the fastest technique, generating minimized test suites in less than a second. 
	However, as discussed in Section \ref{RQ1}, Nemo suffers from a significant loss in statement coverage and fault detection ability. Therefore, although it can provide solutions quickly, the reduced test suites are not effective, rendering it impractical. 
	\toolname consistently provides effective solutions, in terms of coverage and fault finding, for all subjects in under 47 minutes. \toolname outperforms \textsc{TripLEX} and ATM, especially for larger test suites, where \textsc{TripLEX} fails due to timeouts and memory issues, and ATM takes over 10 hours for larger  subjects~\cite{pan2023atm} since its algorithm, irrespective of the similarity measurement, scales quadratically with the number of test cases. 
	
	To solve our ILP formulation, users have the flexibility to utilize commercial solvers like CPLEX directly for small subjects containing fewer than 1,000 test cases. However, for scaling to larger test suites, our RL approach \toolname emerges as the superior choice for addressing the MCTSM problem in practice, especially in terms of both time efficiency and reduced test suite effectiveness.  
	
	\toolname runtime shows a linear relationship with test suite size, production code size and TSM instance size, which demonstrates scalability compared to other techniques. Even when scaling up to test suites of 10,000 tests, 100,000 lines of production code and TSM instance sizes of 10 million edges, \toolname can efficiently reduce the test suite in approximately 3 hours, showcasing its robust scalability and efficiency for large-scale applications in practice.

	\noindent\textbf{\toolname vs.\ \textsc{TripLEX}.} Both \toolname and \textsc{TripLEX} follow the ILP formulation described in Section \ref{ilp_form}. The key difference is that \toolname uses RL to solve the ILP formulation, while \textsc{TripLEX} directly utilizes the commercial ILP solver CPLEX. We designed the RL approach to address the scalability issues that ILP solvers encounter with large-scale test minimization problems. The results in Section \ref{time} demonstrate that CPLEX is unworkable for subjects with test suites larger than 3,837, due to either timeout or out-of-memory errors. This confirms that commercial ILP solvers like CPLEX (used by \textsc{TripLEX}) are not suitable for solving large-scale test minimization problems, which perfectly aligns with our design goals for \toolname.
	To ensure consistency in evaluation, we measured \toolname's runtime for the scenario with 10,000 RL training steps. However, we observed rapid convergence of RL training for small subjects, suggesting the potential for early termination of training before reaching the full 10,000 steps. In practice, users have the flexibility to design different strategies tailored to their specific requirements. Parameters such as the duration of RL training, total steps of training, and the number of vectorized environments can be adjusted according to user preferences, as specified in our open-sourced implementation \cite{repo}. Additionally, our implementation includes an early stopping feature, allowing training to terminate before reaching the \textit{total\_timesteps} limit based on certain conditions. For example, monitoring the value function enables stopping training if there has been no improvement over the last 20 evaluations. Therefore, \toolname can be more efficient and competitive compared to \textsc{TripLEX} for smaller subjects in practical settings. Moreover, users have the option to use \toolname's output as a warm start for the ILP solver CPLEX to further optimize the results. 
	
	\noindent\textbf{Threats to Validity.} The primary concern regarding internal validity is the potential presence of implementation bugs that could affect our results. To mitigate this risk, we compare several popular RL libraries \cite{SpinningUp2018, stable-baselines3, shengyi2022the37implementation} to carefully select the algorithm implementation that aligns with our research objectives. Additionally, we check our customized environment implementation to follow the gym API \cite{towers_gymnasium_2023}. Using a limited number of subject systems in our evaluation poses an external validity of our results. We tried to minimize the threat to select real subjects from the popular dataset used in software engineering research \cite{just2014defects4j}. Regarding replicability, our implementation and empirical data is available online \cite{repo}.
	
	\section{Related Work}\label{related}
	Various techniques \cite{yoo2012regression, khan2018systematic} have been proposed to address test minimization problem. Notably, Integer Programming (IP) has been a valuable tool for formulating and solving Multi-Criteria Test-Suite Minimization (MCTSM) problems for nearly two decades. Black et al. \cite{black2004bi} formulate the problem as a bi-criteria ILP model, taking into account the ability of individual test cases to reveal errors. Hsu and Orso \cite{hsu2009mints} propose the MINTS framework, which extends and generalizes the technique proposed by Black et al. \cite{black2004bi}. MINTS is able to accommodate an arbitrary number of objectives and constraints, making it a flexible approach to formulate MCTSM problems. Hao et al. \cite{hao2012demand} focus on controlling the fault-detection loss of the reduced test suite by collecting statistics on statement-level fault-detection loss into ILP constraints. Additionally, Baller et al. \cite{baller2014multi} apply ILP to solve the test-suite minimization (TSM) problem for the sets of software variants under test. They employ a weighted summation policy to handle multiple conflicting optimization objectives. Lin et al. \cite{lin2018nemo} point out that the previous ILP models \cite{black2004bi, hsu2009mints} are not sound because they did not consider the dependency between test cases and faults. Hence, they introduce Integer Nonlinear Programming (INP) to address this issue and then convert it into a linear model with extra decisive variables. Recently, Xue et al. \cite{xue2020multi} applied Multi-Objective Integer Programming methods for the TSM problem without using a weighted summation policy for various optimization criteria. They also controvert the conclusion of Lin et al. \cite{lin2018nemo} and state that the TSM problem can be adequately modeled with ILP without the need for an INP formulation. Prior research on MCTSM primarily focuses on formulating the TSM problem and solving it directly using commercial ILP solvers. However, the computational complexity of the TSM problem, classified as NP-hard, renders the ILP solvers impractical for large or industrial-scale test suites. 
	
	In addition to Multi-Criteria techniques, which extracts coverage between production and test code, black-box test minimization techniques \cite{cruciani2019scalable,pan2023atm} rely solely on test code with the aim of prioritizing computational speed, they may compromise the fault detection rate of the test suite \cite{cruciani2019scalable} and require the user to specify a budget representing the desired size of the reduced test suite. Furthermore, empirical evidence in \cite{pan2023atm} shows that these approaches still struggle with scalability when confronted with large test suites, posing potential limitations in detecting critical issues within large-scale codebases. Distinguishing ourselves from prior research, we pioneer the integration of coverage diversity of the reduced test suite into the ILP formulation, aiming to enhance the capability to detect unknown faults. Additionally, we introduce a more scalable approach to resolve the minimization problem for large test suites in real-world settings by leveraging Reinforcement Learning.
	
	Over the past few years, researchers in areas such as optimization and operations have increasingly turned their attention to employing Reinforcement Learning (RL) techniques for tackling NP-hard combinatorial optimization problems.  \cite{nie2023reinforcement, mazyavkina2021reinforcement}. This shift reflects the potential of RL to provide innovative and scalable solutions to complex, computationally demanding problems. Khalil et al. \cite{khalil2017learning} introduce the S2V-DQN framework, a pioneering combination of Reinforcement Learning (RL) and Graph Embedding designed to address a range of popular combinatorial optimization problems on graphs. Despite its innovative approach, the framework encounters scalability limitations, notably struggling with graphs exceeding 20,000 nodes. Manchanda et al. \cite{manchanda2019learning} target the scalability challenge posed by combinatorial problems on billion-sized graphs. They address it by training a Graph Convolutional Network (GCN) and integrating it with a Q-learning framework. Their approach demonstrates remarkable speed improvements, being 100 times faster while maintaining solution quality, compared to other algorithms for solving large-scale combinatorial problems. Zhu et al. \cite{zhu2021network} propose a deep reinforcement learning approach for large-scale network planning problems. They also leverage GCN to encode the network's topology and generate graph embeddings. The integration of RL with Integer Linear Programming (ILP) represents a key feature of their approach, effectively addressing the scalability challenges that arise in the context of large-scale network topologies. In contrast, our focus lies in tackling the test minimization problem within the software engineering domain. Rather than relying on graph neural networks for graph embeddings, we opt for a more lightweight yet precise bipartite graph embedding algorithm \cite{yang2022scalable}. This choice is made specifically to align with our MCTSM problem, avoiding the need for repetitive iterations of GCN. Furthermore, we meticulously design our RL training using the Proximal Policy Optimization (PPO) algorithm, tailoring it to suit the unique characteristics of our problem.
	
	While several existing techniques have explored the application of RL to address the test prioritization and test selection problems~\cite{spieker2017reinforcement, bagherzadeh2021reinforcement, bertolino2020learning}, our work stands out as the first, to the best of our knowledge, to employ RL in the test suite minimization problem. Notably, we not only formulate the MCTSM problem as an ILP model but also represent it as a bipartite graph. To enhance the RL-based learning process, we incorporate state-of-the-art bipartite graph embedding techniques. Our proposed approach effectively addresses the MCTSM problem by combining Reinforcement Learning with ILP techniques, holds significant promise for minimizing test suites in a scalable manner.

	\section{Conclusions}\label{conclusion}
	The Multi-Criteria Test Suite Minimization (MCTSM) problem aims to refine test suites by removing redundant test cases based on metrics like code coverage and fault detection. Traditional methods use integer linear programming (ILP) but face scalability challenges due to the NP-hard nature of MCTSM.
	In this work, we introduced a novel strategy that incorporates statement coverage, fault detection constraints, and minimizes test suite size and pairwise coverage similarity into the ILP formulation. We also leverage Reinforcement Learning (RL) to solve this ILP model, enhancing efficiency and scalability for large test suites. Our empirical evaluation shows that our approach, \toolname, scales linearly with problem size, delivering solutions in less than 47 minutes for larger programs from Defects4J where ILP solvers struggle. \toolname also outperforms related techniques in test effectiveness.
	Looking ahead, we aim to deploy our strategy on more real-world test suites and fine-tune hyperparameters for further performance improvements.
	\nocite{*}
	\bibliographystyle{ACM-Reference-Format}
	\bibliography{main}
	
\end{document}